\documentclass[12pt]{article}
\usepackage{amsmath,amssymb,bbold,hyperref,graphicx,color,soul,xcolor,physics,bm,mathrsfs}
\newcommand{\be}{\begin{equation}}
\newcommand{\ee}{\end{equation}}
\newcommand{\bea}{\begin{eqnarray}}
\newcommand{\eea}{\end{eqnarray}}
\newcommand{\beas}{\begin{eqnarray*}}
\newcommand{\eeas}{\end{eqnarray*}}

\newcommand{\Oi}{\mathcal{O}_i}
\newcommand{\Oj}{\mathcal{O}_j}

\def\be{\begin{eqnarray}}
	\def\ee{\end{eqnarray}}

\def\>{\rangle}
\def\<{\langle}
\def\lb{\label}
\def\zb{\bar{z}}
\def\wb{\bar{w}}
\def\tb{\bar{T}}
\def\bb{\bar{B}}
\def\lb{\bar{L}}
\def\rb{\bar{R}}

\def\ba{\begin{align}}
	\def\ea{\end{align}}
\def\bas{\begin{align*}}
	\def\eas{\end{align*}}

\begin{document}
\begin{titlepage}

\begin{center}

{\Large Revisiting subregion holography using OPE blocks}

\vspace{12mm}

\renewcommand\thefootnote{\mbox{$\fnsymbol{footnote}$}}
Mrityunjay Nath${}^{1}$\footnote{nath.mrityunjay@gmail.com},
Satyabrata Sahoo${}^{1}$\footnote{satya101@live.com} and
Debajyoti Sarkar${}^{1,2}$\footnote{dsarkar@iiti.ac.in}

\vspace{6mm}

\vspace{2mm}
${}^1${\small \sl Department of Physics} \\
{\small \sl Indian Institute of Technology Indore} \\
{\small \sl Khandwa Road 453552 Indore, India}

\vspace{4mm}
${}^2${\small \sl Institute for Advanced Study } \\
{\small \sl Tsinghua University} \\
{\small \sl Beijing 100084, China.}

\end{center}

\vspace{12mm}

\noindent
In this short note, we revisit the entanglement wedge representation of AdS$_3$ bulk fields in terms of CFT operator product expansion (OPE) blocks for a general class of blocks. Given a boundary interval and its associated causal diamond, the OPEs involve boundary operators with or without spin, and located either at spacelike or timelike edges of the diamond. Only for a subset of these cases, can the OPE block be dual to a geodesic bulk field. We show that when applied to de Sitter, a suitable combination of Euclidean OPE blocks can represent a dS scalar integrated over the timelike extremal surfaces, which play an important role in defining pseudo-entropy. We also work out some simple higher dimensional examples. 

\end{titlepage}
\setcounter{footnote}{0}
\renewcommand\thefootnote{\mbox{\arabic{footnote}}}

\hrule
\tableofcontents
\bigskip
\hrule

\addtolength{\parskip}{8pt}
\section{Introduction}\label{sec:introduction}

Operator product expansion (OPE) encapsulates the operator spectrum of the field theory along with the corresponding data of its OPE coefficients. For a conformal field theory (CFT), given two scalar quasi-primaries $\Oi$ and $\Oj$ of conformal dimensions $\Delta_i$ and $\Delta_j$, we can write down their OPE as 
\begin{equation}
	\Oi(x)\Oj(0)=\sum_k C_{ijk}\,|x|^{\Delta_k-\Delta_i-\Delta_j}\left(1+c_1x^\mu\partial_\mu+c_2x^\mu x^\nu\partial_\mu\partial_\nu+\dots\right)\mathcal{O}_k(0)\,,
\end{equation}
where $C_{ijk}$ are the OPE coefficients and $O_k$ are the channel operators that appear in the
OPE. More generally, such an OPE can be repackaged in terms of the OPE block $\mathcal{B}_{\Delta_k,(\ell_1+\ell_2)}^{ij}$ in the operator channel $\mathcal{O}_k$ as
\begin{equation}
	\mathcal{O}_{\Delta_i,\ell_1}(x_1)\mathcal{O}_{\Delta_j,\ell_2}(x_2)=|x_1-x_2|^{-\Delta_i-\Delta_j}\sum_k C_{ijk}\,\mathcal{B}_{\Delta_k,(\ell_1+\ell_2)}^{ij}\,.
\end{equation}
We have written the expansion above for the case when the quasi-primaries are allowed to have non-zero spins $\ell_i$'s. In a holographic setup, OPE blocks were introduced in \cite{Czech:2016xec,deBoer:2016pqk} which were then used to study the boundary to bulk maps in Anti de Sitter (AdS)/CFT dictionary. 
Moreover, the connection between certain OPE blocks and modular Hamiltonians (as was addressed in the original papers above) can then be used to probe the bulk and boundary modular flows, which plays a pivotal role in the study of quantum entanglement. In this paper we will revisit these old ideas and close a few literature gaps in their applicability towards bulk reconstruction.

One of the main goals of AdS/CFT correspondence is to understand how the bulk geometry and dynamics emerge from the boundary conformal field theory. A key tool in this endeavor is the CFT extrapolate dictionary, which allows us to reconstruct local bulk fields from CFT operators and correlators. The original version of this dictionary, proposed in \cite{Banks:1998dd,Bena:1999jv,Balasubramanian:1998sn,Balasubramanian:1999ri}, required the knowledge of the CFT operators on an entire Cauchy slice, rather than a partial set of information confined within a subregion. A significant improvement was achieved in \cite{Hamilton:2006az,Hamilton:2006fh,Kabat:2011rz,Kabat:2012hp,Kabat:2012av,Kabat:2013wga}, where the authors developed a revised version of the dictionary which can be easily extended to the so-called subregion holography.\footnote{Based on the original authors, we will call this revised approach HKLL prescription.} We turn to this topic next. 

A refined approach towards the extrapolate dictionary started with the advent of Ryu-Takayanagi (RT) \cite{Ryu:2006bv} and covariant HRT \cite{Hubeny:2007xt} prescriptions, which conjectured that the entanglement entropy of a boundary subregion is given by the area of a minimal/extremal surface in the bulk that anchors on the boundary of the subregion. This establishes a holographic relation between a subregion of the boundary and a corresponding region in the bulk, called the entanglement wedge (EW) reconstruction \cite{Faulkner:2017vdd,Dong:2016eik}. This approach incorporates quantum information theoretic concepts such as relative entropy and modular Hamiltonian into holography, and reveals new aspects of bulk emergence from boundary entanglement.\footnote{The applications of relative entropy and the modular Hamiltonian in the holographic context has a long history. See e.g. \cite{Casini:2011kv,Blanco:2013joa} to name a few.}

For completeness, we wish to recollect the main features and applications of the CFT extrapolate dictionary in Lorentzian AdS/CFT correspondence, with a focus on subregion holography. The original HKLL approach, which relied on solving bulk equations of motion, provided a position space expression of the bulk field $\phi(x,z)$ dual to a boundary operator $\mathcal{O}_{\Delta}(x)$
\begin{equation}
	\phi(z,x)=\int dx'K_\Delta(z,x|x')\mathcal{O}_\Delta(x')+\sum_{l}\frac{a_l}{N}\int dx' K_{\Delta_l}(z,x|x')\mathcal{O}_{\Delta_l}(x')+\mathcal{O}(N^{-2})\dots\,.
\end{equation}
Here $z$ is the bulk radial direction, $x$ are boundary coordinates, $K$ is the so-called smearing Kernel. The expression above provides an order by order construction of the bulk field in the boundary effective coupling $1/N$ of a CFT of large central charge $\sim N$. The information of bulk interaction is encoded via the appropriate coefficients $a_l$. One can use this prescription for a bulk field in any coordinate patch such as in Rindler spacetimes \cite{Hamilton:2006az,Hamilton:2006fh}, or for that matter, inside any bulk EW region dual to a corresponding finite subregion at the boundary. In these cases, the bulk field is obtained via boundary operators supported over an \emph{imaginary} spatial direction $i\rho'$, and we consider a `spacelike' relation between this $\rho'$, time $\chi'$ and bulk radial coordinate $\tilde{r}$ (for a description of these coordinates, see appendix \ref{appendix:matrices}. Also see figure \ref{fig:rindler}) over which the smearing kernel is supported. For example, for a free bulk scalar $\phi$ located outside the AdS$_3$ Rindler horizon (corresponding to a boundary scalar primary $\mathcal{O}$ of conformal dimension $\Delta$) we have
\begin{equation}
	\phi_{Rindler}(\tilde{r},\chi,\rho)=C_{\Delta}\int_{\sigma>0} d\rho'd\chi'\, \left(\lim_{\tilde{r}'\to \infty}K(\tilde{r},\chi,\rho|\tilde{r}',\chi+\chi',\rho+i\rho')\right) \,\mathcal{O}_{\Delta}(\chi+\chi',\rho+i\rho').
\end{equation}
Here $C_{\Delta}$ is an overall normalization ($C_\Delta=\frac{(\Delta-1)2^{\Delta-2}}{\text{Vol}(\mathrm{B}^2)}$, with $\mathrm{B}^2$ being a 2-ball of unit radius) and $K$ is the smearing kernel which is an appropriate function of AdS bulk-boundary covariant distance. It is related to the bulk to bulk covariant distance $\sigma$ as \cite{Hamilton:2006fh}
\begin{equation}
	\lim_{\tilde{r}'\to \infty}K(\tilde{r},\chi,\rho|\tilde{r}',\chi+\chi',\rho+i\rho')=\lim_{\tilde{r}'\to \infty}\left(\sigma/\tilde{r}'\right)^{\Delta-2}=\left[\tilde{r}\left(\cos \rho'-\left(1-\tilde{r}^{-2}\right)^{1/2}\cosh\chi'\right)\right]^{\Delta-2}.
\end{equation}
At this stage, let us make some comments on this above-mentioned complexification. For a position space bulk reconstruction in Rindler, such complexifications become a necessity and can be traced back to the presence of evanescent modes for bulk fields in a background containing a horizon \cite{Bousso:2012mh,Leichenauer:2013kaa,Rey:2014dpa}. As a result, the HKLL smearing kernel should be understood as a distribution in position space \cite{Morrison:2014jha}. Similar games can be played for global and Poincar\'{e} coordinates, but there it is essentially a useful trick. An alternate way to think about this procedure is via identifying the HKLL kernel as a retarded Green's function in de Sitter (dS) \cite{Hamilton:2006az,Hamilton:2006fh}. But it is not clear if this dS interpretation should go through for bulk reconstructions in any excited state. Because CFT is considered to be the input in this approach, such complexification is indeed mysterious solely from the boundary perspective. So far, any interpretation or explanation of this complexification involves having an idea of the associated bulk geometry, particularly the existence of the horizon. 

On the other hand, it was shown in  \cite{Czech:2016xec,deBoer:2016pqk} that in \emph{certain examples}, the OPE block repackages the bulk fields as an inverse geodesic Radon transform (or X-ray transform)  with a given weight (there is also a connection with Mellin transform; see e.g. \cite{daCunha:2016crm,Guica:2016pid}. See also \cite{Bhowmick:2018bmn,Bhowmick:2019nso}). In particular, in AdS$_3$/CFT$_2$, if the OPE block has been produced by two \emph{identical} boundary scalar operators located at the endpoints of a given RT surface $\gamma$, it provides a representation of a bulk field $\phi$ integrated over the bulk RT surface (parametrized by $s$ with $c_\Delta$ being a $\Delta$ dependent constant)\footnote{Throughout this paper, we will only consider free fields. Interacting fields require a knowledge of bulk interactions, and will relate to $1/N$ corrections to boundary OPE blocks. The inclusion of interactions is quite orthogonal to the topics we are addressing in this paper.}
\begin{equation}
	\int_{\gamma} ds\, \phi(s)=c_\Delta \mathcal{B}_k+\mathcal{O}(N^{-1})\dots\,.
\end{equation}
We should note (as it will be useful for us later) that purely within field theory, OPE blocks can also be defined in terms of the shadow operator prescription in the CFT. For a boundary interval $[L,R]$, it takes the form 
\begin{align}
	\mathcal{B}^{k}_{(\Delta)}(L,R)&=\tilde{c}_\Delta\int_{-\infty}^{\infty}\int_{-\infty}^{\infty}d\chi\, d\rho\,\left\langle\mathcal{O}_{L}(0,-\infty)\mathcal{O}_{R}(0,\infty)\tilde{\mathcal{O}}_{\tilde{\Delta}}(\chi,\rho)\right\rangle\mathcal{O}_{\Delta}(\chi,\rho)\,,
\end{align}
where the shadow operator $\tilde{O}$ has dimension $\tilde{\Delta}=d-\Delta$ \cite{SimmonsDuffin:2012uy,Czech:2016xec,deBoer:2016pqk}. Throughout this paper, we will refer the operators $\Oi$ and $\Oj$, participating in the OPE, as \emph{external} operators.

A good part of our work will be devoted to showing that the above observation is a special example of a much broader picture. In particular, in two-dimensions, we will derive a dictionary between geodesic bulk fields and OPE blocks, where the external operator in the OPE could in general be spinning and can also be time-like separated. Whereas, in higher dimensions, we will show that such a connection is unavailable.\footnote{Considerations of such top-bottom OPE blocks is not new, and was mentioned earlier in \cite{Czech:2016xec,deBoer:2016pqk,Chen:2019fvi}. However, especially in \cite{Czech:2016xec,deBoer:2016pqk} the discussion was heuristic, and such OPE blocks were discussed using surface OPE and kinematic duality arguments, and in terms of surface Witten diagrams. Although our results for $d=2$ match with those, we find some crucial subtleties in higher dimensions that were not emphasized in those studies.} 

Note that, whenever we have a boundary entangling region in the context of subregion duality, we have a natural causal diamond (CD) associated with it. We can therefore consider the external operators to be located either at the spacelike endpoints of the entangling region (distinct points for CFT$_2$), or at the timelike separated top and bottom points of the CD. Analyzing some generic cases, we will then find that possibly a better-phrased equivalence between the HKLL fields and OPE blocks comes in terms of Fourier transformation. This is especially true, if motivated from the dS/CFT extrapolate dictionary, we consider the OPE blocks between operators from a principal series representation \cite{Kravchuk:2018htv,Chatterjee:2016ifv,Gadde:2017sjg}. We will see that our statement above naturally justifies the complexification mentioned earlier, by means of an analytic continuation from \emph{dS to AdS}. As we'll show, this has the added feature of being able to interpret bulk fields in dS spacetime as a superposition of Euclidean OPE blocks. In these cases the OPE blocks become associated to the bulk fields integrated over timelike extremal surfaces, which has been a topic of recent interests in the study of timelike entanglement entropy within holography \cite{Nakata:2020luh,Doi:2022iyj,Narayan:2022afv,Doi:2023zaf,Narayan:2023zen}.

For the convenience of the reader, here we will briefly summarize the structure and the main results of our paper. The first few sections will only deal with the AdS$_3$/CFT$_2$ correspondence, and only near the end we will discuss higher dimensional generalizations. 
\begin{itemize}
	\item In section \ref{sec:RHKLL} we start by reviewing the HKLL bulk reconstruction for AdS$_3$ Rindler, and recall its connection with OPE blocks of two dimensional CFT in section \ref{sec:ope}. Even though many of these results are quite well-known and should be thought of as a necessary review, we have computed some newer types of OPE blocks in section \ref{sec:ope} where the external operators could be timelike separated and with higher spins. See e.g.~equations \eqref{eq:opeeqk4} and \eqref{eq:opeeqk5}. We will see that in some of these cases, they also have the interpretation of smearing AdS$_3$ bulk scalars along the RT geodesic. These conclusions will be drawn in section \ref{sec:opehkllmatch}.
	\item The analysis of the above sections motivates us to ask analogous questions in the de Sitter context. This has been carried out in section \ref{sec:dSconn}. Here, we write down the bulk fields integrated over the spacelike hyperbolas  as a combination of Euclidean OPE blocks. We have performed this in the flat slicings of dS. It is to be noted that these spacelike hyperbolas appear in the context of de Sitter pseudo-entropy, and have also recently appeared as part of extremal surfaces for a timelike subregion in AdS.
	\item In section \ref{sec:higherd}, we then turn to the higher dimensional OPE blocks and make some observations regarding their possible connections with  higher dimensional bulk fields. Some of the important expressions there are \eqref{eq:BJOO3d} through \eqref{eq:zbzbtopebhshd}. The entire analysis in these main parts of the paper will be for bulk scalar fields.
\end{itemize}

We have provided some appendices at the end to supplement the readers with the necessary backgrounds and analytic justifications. In particular, appendix \ref{appendix:matrices} introduces some notations, coordinate patches, basics of modular flow etc.~which we have utilized throughout the paper. Appendix \ref{app:HBcomm} provides the connection between modular Hamiltonians, modular momentums and OPE blocks in two and higher dimensional CFTs. This plays an essential role in order to understand their relation with the free bulk fields. Appendix \ref{app:spin2case} is entirely devoted to the bulk graviton reconstruction in the context of AdS$_3$/CFT$_2$. And finally appendix \ref{sec:OPEdS} supplements some of the details of the calculations carried out in section \ref{sec:dSconn}.

\section{Free scalars in AdS$_3$ Rindler}\label{sec:RHKLL}

This section will revisit the HKLL extrapolate dictionary for the free bulk scalars. Barring some differences, much of the analysis presented here already appears in \cite{Kabat:2017mun}. Therefore, we will be very brief.

As mentioned in the introduction, the Rindler dictionary takes the form
\begin{equation}\label{eq:RHKLLinR}
	\phi_{Rindler}(\tilde{r},\chi,\rho)=C_{\Delta}\int_{\sigma>0} d\rho'd\chi'\, \left(\lim_{\tilde{r}'\to \infty}K(\tilde{r},\chi,\rho|\tilde{r}',\chi+\chi',\rho+i\rho')\right) \,\mathcal{O}_{\Delta}(\chi+\chi',\rho+i\rho')\,,
\end{equation}
with the smearing kernel 
\begin{equation}\label{eq:KRHKLL}
	\lim_{\tilde{r}'\to \infty}K(\tilde{r},\chi,\rho|\tilde{r}',\chi+\chi',\rho+i\rho')=\lim_{\tilde{r}'\to \infty}\left(\sigma/\tilde{r}'\right)^{\Delta-2}=\left[\tilde{r}\left(\cos \rho'-\left(1-\tilde{r}^{-2}\right)^{1/2}\cosh\chi'\right)\right]^{\Delta-2}.
\end{equation}
\begin{figure}[t]
	\begin{center}
		\includegraphics[width=85mm]{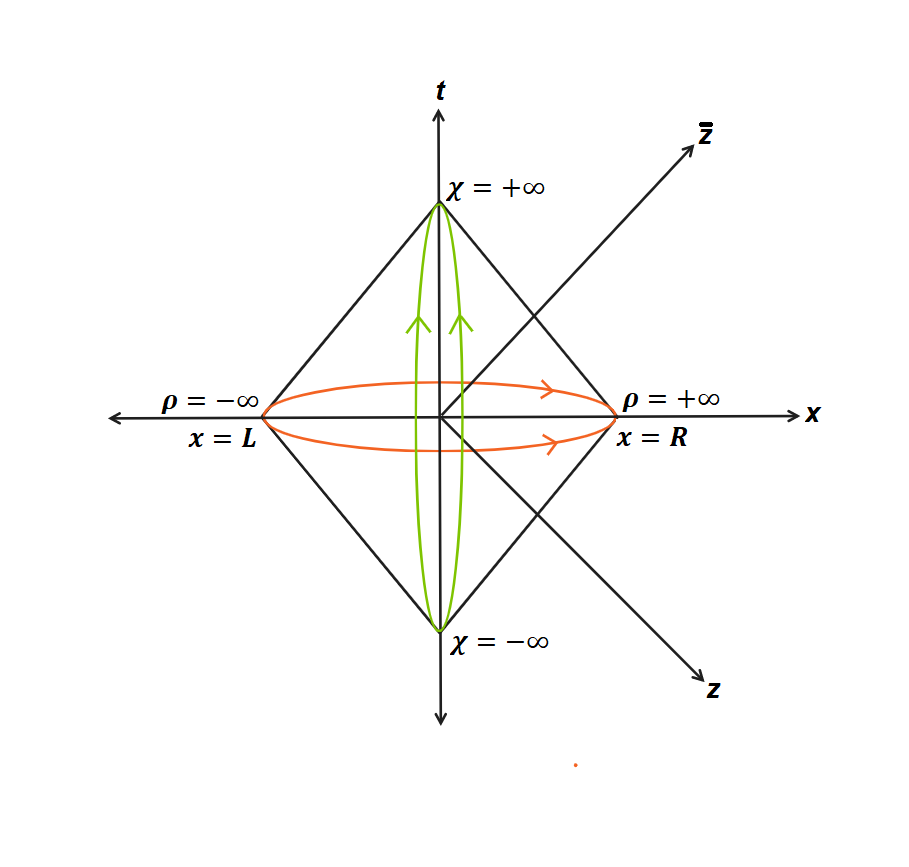}
		\caption{Causal diamond of a spatial subregion (on a CFT$_2$ plane) in both lightcone $(z,\zb)$ and Rindler coordinates $(\rho,\chi)$. Rindler coordinates span the entire causal diamond. The green and red arrows signify modular flow directions by modular Hamiltonian $H_{mod}$ and modular momentum $P_{mod}$ respectively. See appendix \ref{appendix:matrices} for detailed definitions of these quantities.}
		\label{fig:rindler}
	\end{center}
\end{figure}
Quite interestingly, given the $H_{mod}$ and $P_{mod}$ make a boundary operator flow in the $\chi$ and $\rho$ directions respectively (see appendix \ref{appendix:matrices}), the above boundary to bulk map can also be alternatively represented as 
\begin{align}\label{eq:RHKLLinR2}
	\phi_{Rindler}(\tilde{r},\chi,\rho)&=C_{\Delta}\int_{\sigma>0} d\rho'd\chi'\, \left[\tilde{r}\left(\cos \rho'-\left(1-\tilde{r}^{-2}\right)^{1/2}\cosh\chi'\right)\right]^{\Delta-2}\nonumber\\
	&e^{-\frac{i\left(P_{mod}\,(i\rho')+H_{mod}{\chi'}\right)}{2\pi}}\mathcal{O}_{\Delta}(\chi,\rho)e^{\frac{i\left(H_{mod}{\chi'}+P_{mod}\,(i\rho')\right)}{2\pi}}\,.
\end{align}
In other words, the spatial complexification of $\mathcal{O}_{\Delta}$ can be understood as a flow under $P_{mod}$ by a certain imaginary value. The integration region $\sigma>0$ constraint implies
\begin{equation}
	\cos\rho'>\left(1-\tilde{r}^{-2}\right)^{1/2}\cosh\chi'\,.
\end{equation}
The above result simplifies considerably for when the bulk operator is on the RT surface $\tilde{r}=1$, which gives the integration region as $\cos{\rho'}>0$.
In this case, the ${\chi'}$ integral runs from $\{-\infty,\infty\}$, whereas the ${\rho'}$ integral runs between $\{-\pi/2,\pi/2\}$.\footnote{Note that as $\rho'$ is real, we can think of the flow by $P_{mod}$ as a flow by an imaginary parameter $s_{\rho}=i\rho'$. In modular theory there is a regime of validity for the imaginary modular flow parameter \cite{Witten:2018zxz}.
	When the bulk field is at the RT surface, i.e. $s_\rho=\{-i\pi/2,i\pi/2\}$, this regime of validity is just saturated. Hence interpreting the complexification of boundary spatial coordinates in this manner is well-defined for all regions inside the entanglement wedge.} 
We therefore obtain
\begin{align}\label{eq:simp1}
	&\phi_{Rindler}(\tilde{r}=1,\chi,\rho)\nonumber\\
	&=C_{\Delta}\int_{-\pi/2}^{\pi/2} d{\rho'}\int_{-\infty}^{\infty}d{\chi'}\, \left(\cos \rho'\right)^{\Delta-2}e^{-\frac{i\left(P_{mod}\,(i\rho')+H_{mod}{\chi'}\right)}{2\pi}}\mathcal{O}_{\Delta}(\chi,\rho)\,e^{\frac{i\left(H_{mod}{\chi'}+P_{mod}\,(i\rho')\right)}{2\pi}}\nonumber\\
	&=C_{\Delta}\int_{-\pi/2}^{\pi/2} d{\rho'}\, \left(\cos\rho'\right)^{\Delta-2}e^{-\frac{iP_{mod}\,(i\rho')}{2\pi}}
	\left( \int_{-\infty}^{\infty}ds\, \mathcal{O}_{\Delta}(s,\rho) \right)\, e^{\frac{iP_{mod}\,(i\rho')}{2\pi}}\,.
\end{align}
Note that in going from the second line to the last line above, we've changed the variable of integration $\chi' \to s = \chi + \chi'$. For now we have kept the $\chi$ on the left hand side (as an argument of the bulk field), but from the right hand side, the resulting quantity is clearly $\chi$ independent. This is of course expected as $\chi$ is a hyperbolic angle relative to the RT surface and it ceases to be well-defined as we place a bulk field on the RT surface. In what follows, we will therefore avoid putting $\chi$ in the argument of the bulk field, as they will always be located on the RT surface.

Using the following convention for Fourier transform 
\begin{equation}\label{eq:ftomega}
	\mathcal{O}_{\Delta}({\chi'},\rho)=\int_{-\infty}^{\infty}d\omega_{\chi'}\,e^{-i\omega_{\chi'}{\chi'}}\mathcal{O}_{\Delta}(\omega_{\chi'},\rho),
\end{equation}
and plugging it in \eqref{eq:simp1} above, we see that it puts $\omega_{\chi'}=0$. We therefore have
\begin{equation}\label{eq:simp2}
	\phi_{Rindler}(\tilde{r}=1,\rho)=2\pi C_{\Delta}\int_{-\pi/2}^{\pi/2} d{\rho'}\, \left(\cos \rho'\right)^{\Delta-2}e^{-\frac{iP_{mod}\,(i\rho')}{2\pi}}\mathcal{O}_{\Delta}(\omega_{\chi'}=0,\rho)e^{\frac{iP_{mod}\,(i\rho')}{2\pi}}.
\end{equation}
Fourier transforming in a similar manner in the $\rho'$ direction, we end up with\footnote{Same result is obtained by expanding $e^{-\frac{iP_{mod}\,(i\rho')}{2\pi}}\mathcal{O}_{\Delta}(\omega_{\chi'}=0,k_{\rho})\,e^{\frac{iP_{mod}\,(i\rho')}{2\pi}}$ in \eqref{eq:simp2}, and taking $P_{mod}=i\partial_\rho$.}
\begin{equation}
	\phi_{Rindler}(\tilde{r}=1,\rho)=2\pi C_{\Delta}\int_{-\infty}^{\infty}dk_{\rho}\,e^{ik_{\rho}\rho}\left(\frac{\Gamma(\Delta)}{2^{\Delta-1}(\Delta-1)|\Gamma\left(\frac{\Delta}{2}+\frac{ik_\rho}{2}\right)|^2}\right)\,\mathcal{O}_{\Delta}(\omega_{\chi'}=0,k_{\rho})\,.
\end{equation}
Inverting the above relation, we finally obtain
\begin{equation}\label{eq:fttransformedmap-new}
	\mathcal{O}_{\Delta}(\omega_{\chi'}=0,k_{\rho})=\left(\frac{2^{\Delta-1}(\Delta-1)|\Gamma\left(\frac{\Delta}{2}+\frac{ik_\rho}{2}\right)|^2}{4\pi^2C_{\Delta}\Gamma(\Delta)}\right)\int_{-\infty}^{\infty}d\rho\,e^{-ik_{\rho}\rho}\,\phi_{Rindler}(\tilde{r}=1,\rho)\,.
\end{equation}
This is a known result which shows that the zero (energy) mode of the boundary primary is a weighted integral (Fourier transform) of the bulk field over the RT surface.

\section{Various OPE blocks in CFT$_2$}\label{sec:ope}

In this section, we switch our attention to OPE blocks in CFT$_2$. In order to connect with the Rindler HKLL prescription for free bulk scalars, we can consider \emph{external} boundary CFT quasi-primary operators located at either the left-right or top-bottom points of the boundary causal diamond $D$, which is associated  to the boundary subregion at $t=0$ slice. The external operators will be taken to have unequal spins and conformal dimensions (so, an arbitrary twist in general), which we need to study in the scalar channel. We will explore these various scenarios in subsections \ref{subsec:scalarope} and \ref{subsec:phiTope}. Later on, in subsections \ref{subsec:RHKLLgrav} and \ref{subsec:OPEisH}, we write down the OPE blocks in stress-tensor channel in order to connect with the bulk graviton reconstruction. 

\subsection{Scalar channel OPE blocks between scalar primaries}\label{subsec:scalarope}

\emph{Left-right blocks:}
We start by studying the CFT$_2$ OPE block $\mathcal{B}_{\mathcal{O}\mathcal{O}\mathcal{O}_\Delta}(L,R)$ in scalar $\mathcal{O}_\Delta$ channel. These particular blocks were studied in detail in \cite{daCunha:2016crm,Czech:2016xec,deBoer:2016pqk}. The external operators here have conformal dimensions $\Delta_L$ and $\Delta_R$ and the subscripts $L$ and $R$ designate that the operators are located at left and right endpoints of the subregion respectively. We analyze these OPE blocks in the shadow operator formalism given in e.g. \cite{Ferrara:1972uq,Ferrara:1972ay,SimmonsDuffin:2012uy}. Once again using the notations of appendix \ref{appendix:matrices} (\eqref{eq:lrinz}-\eqref{eq:cttb}), we have 
\begin{align}\label{eq:opeb}
	&\mathcal{B}_{\mathcal{O}\mathcal{O}\mathcal{O}_\Delta}(L,R)\nonumber\\
	&=C_{\mathcal{O}}\int_{D}dz\,d\zb\,\left\langle\mathcal{O}_{h_L}(L,\bar{L})\mathcal{O}_{h_R}(R,\bar{R})\tilde{\mathcal{O}}_{\tilde{\Delta}}(z,\zb)\right\rangle\mathcal{O}_{\Delta}(z,\zb)\nonumber\\
	&=\tilde{C}_{\mathcal{O}}\int_{D}dz\,d\zb\left(\frac{(z-L)(\zb-\bar{L})}{(R-z)(\bar{R}-\zb)}\right)^{h_R-h_L}\left(\frac{(z-L)(R-z)}{R-L}\right)^{h-1}\left(\frac{(\bar{R}-\zb)(\zb-\bar{L})}{\bar{R}-\bar{L}}\right)^{\bar{h}-1}\mathcal{O}_{\Delta}(z,\zb).
\end{align}
The second line above defines the shadow operator formalism, which includes some overall coefficient $C_{\mathcal{O}}$. In the last line, $h_R=\bar{h}_R=\frac{\Delta_R}{2}$ and $h_L=\bar{h}_L=\frac{\Delta_L}{2}$ are the (holomorphic) conformal dimensions of the operators located at the right and left edges of the RT surface respectively and finally, the operator $\mathcal{O}$ (dual to the free bulk scalar) has a conformal dimension $\Delta=h+\bar{h}$ and spin $\ell=h-\bar{h}$ (which would be zero for bulk scalars).\footnote{There is also a factor $(R-L)^{-(\Delta_L+\Delta_R)}$ present inside the integrand, which we have absorbed in the overall normalization $\tilde{C}_{\mathcal{O}}$. Originally, $\tilde{C}_{\mathcal{O}}$ is given by \cite{Czech:2016xec}
	\begin{equation}\label{eq:COval}
		\tilde{C}_{\mathcal{O}}=\frac{\Gamma(2h)\Gamma(2\bar{h})}{\Gamma(h)\Gamma(\bar{h})}.
	\end{equation}\label{fn:coeff}} 
Rewriting the above equation in Rindler coordinates using \eqref{eq:genctbdry} or \eqref{eq:ct1}, we get (in what follows, $\Delta_{ij}=\Delta_i-\Delta_j$. So, for example, $\Delta_{LR}$ and $\Delta_{TB}$ will always mean the difference in the conformal dimensions between Left-Right or Top-Bottom operators)
\begin{equation}\label{eq:opeeqk}
	\mathcal{B}_{\mathcal{O}\mathcal{O}\mathcal{O}_\Delta}(L,R)=\tilde{C}_{\mathcal{O}}\int_{-\infty}^{\infty}\int_{-\infty}^{\infty}d\chi\, d\rho\, e^{-\frac{\Delta_{LR}}{2}(w+\bar{w})}\mathcal{O}_{\Delta}(\chi,\rho)=\tilde{C}_{\mathcal{O}}\int_{-\infty}^{\infty}\int_{-\infty}^{\infty}d\chi\, d\rho\, e^{-i(-i\Delta_{LR})\,\rho}\mathcal{O}_{\Delta}(\chi,\rho)\,.
\end{equation}
One can check explicitly that $+i\Delta_{LR}$ is precisely the eigenmode of $P_{mod}=H_{mod,rm}-H_{mod,\ell m}$ as defined in \eqref{eq:hmodrhodef}. 
In other words, one can identify $-i\Delta_{LR}$ with $k_{\rho}$, which is the Fourier mode corresponding to $\rho$ if $\Delta_{LR}$ itself is imaginary. This is also substantiated via a direct calculation of how $\mathcal{B}_{\mathcal{O}\mathcal{O}\mathcal{O}_\Delta}(L,R)$ behaves under the modular translation $P_{mod}$. Using the CFT commutator
\begin{equation}\label{eq:tzcomm}
	2\pi [T_{zz}(z),\mathcal{O}]=2\pi i \left[h\,\partial_\xi\delta(\xi-z)\mathcal{O}+\delta(\xi-z)\,\partial_\xi\mathcal{O}\right]
\end{equation}
and its anti-holomorphic counterpart, one indeed finds (we have carried out a detailed analysis of various such commutator relations, including this one, in appendix \ref{app:HBcomm})\footnote{The reason for the minus sign in the middle line of the commutator \eqref{eq:HBcomm} has to do with the fact that $P_{mod}\equiv -i\partial_{\rho}$ brings $-i(-ik_{\rho})$ down from the exponent as the eigenvalue of the OPE block (which is a quantity in the Fourier space).} \cite{Kabat:2017mun,Das:2019iit} \begin{equation}\label{eq:HBcomm}
	\left[P_{mod},\mathcal{B}_{\mathcal{O}\mathcal{O}\mathcal{O}_\Delta}(L,R)\right]=-2\pi\,k_\rho\,\mathcal{B}_{\mathcal{O}\mathcal{O}\mathcal{O}_\Delta}(L,R)=2\pi\,(i\, \Delta_{LR})\,\mathcal{B}_{\mathcal{O}\mathcal{O}\mathcal{O}_\Delta}(L,R)\,.
\end{equation}
As a result, we see that \eqref{eq:opeeqk} is nothing but
\begin{equation}\label{eq:opeeqk2}
	\mathcal{B}_{\mathcal{O}\mathcal{O}\mathcal{O}_\Delta}(L,R)=\tilde{C}_{\mathcal{O}}\int_{-\infty}^{\infty}\int_{-\infty}^{\infty}d\chi\, d\rho\, e^{-ik_{\rho}\,\rho}\mathcal{O}_{\Delta}(\chi,\rho)=4\pi^2 \tilde{C}_{\mathcal{O}}\,\mathcal{O}_{\Delta}(\omega_{\chi}=0,k_{\rho}=-i\Delta_{LR})\,.
\end{equation}
This above subtlety (which disappears if $\Delta_{LR}=0$, which is what's usually taken in the literature for simplicity) sets up one of the observations of our work. From the work of \cite{Gadde:2017sjg,Kravchuk:2018htv,Karateev:2018oml}, it is well known that imaginary conformal dimensions come naturally in the principal series representation of conformal operators. However, here we see that they also arise in the shadow operator prescription of the OPE blocks. In section \ref{sec:dSconn} we will note that this will have a natural place in the connection between the Euclidean OPE blocks and the de Sitter bulk field.

\noindent
\emph{Top-bottom blocks}:
We can also consider the OPE block where the external operators are at a timelike separation, in particular at the top and bottom points of the causal diamond. Such $\mathcal{B}_{\mathcal{O}\mathcal{O}\mathcal{O}_\Delta}(T,B)$ would be a necessary tool for us especially in higher dimensions, where the $L,R$ points are no longer unique points of the boundary causal diamond. Rewriting the shadow operator prescription of \eqref{eq:opeb} in this case, we have
\begin{align}\label{eq:opebtb}
	&\mathcal{B}_{\mathcal{O}\mathcal{O}\mathcal{O}_\Delta}(T,B)\nonumber\\
	&=C_{\mathcal{O}}\int_{D_{z\zb}}dz\,d\zb\,\left\langle\mathcal{O}_{h_T}(T,\bar{T})\mathcal{O}_{h_B}(B,\bar{B})\tilde{\mathcal{O}}_{\tilde{\Delta}}(z,\zb)\right\rangle\mathcal{O}_{\Delta}(z,\zb)\nonumber\\
	&=\tilde{C}_{\mathcal{O}}\int_{D_{z\zb}}dz\,d\zb\left(\frac{(B-z)(\zb-\bar{B})}{(z-T)(\bar{T}-\zb)}\right)^{h_T-h_B}\left(\frac{(z-T)(B-z)}{B-T}\right)^{h-1}\left(\frac{(\bar{T}-\zb)(\zb-\bar{B})}{\bar{T}-\bar{B}}\right)^{\bar{h}-1}\mathcal{O}_{\Delta}(z,\zb).
\end{align}
The notations above have equivalent meaning compared to the left-right case, with the difference that we have now changed the subscripts and arguments to $T,B$ etc.~in order to emphasize that here we are dealing with the top-bottom scenario. We are still using $C_{\mathcal{O}}$ and $\tilde{C}_{\mathcal{O}}$ as constants so as to not introduce a plethora of new symbols, although they should be distinguished from what appeared in \eqref{eq:opeb}. Once again, going to the Rindler coordinates we have
\begin{equation}\label{eq:opeeqk4}
	\mathcal{B}_{\mathcal{O}\mathcal{O}\mathcal{O}_\Delta}(T,B)=\tilde{C}_{\mathcal{O}}\int_{-\infty}^{\infty}\int_{-\infty}^{\infty}d\chi\, d\rho\, e^{-\frac{\Delta_{BT}}{2}(\bar{w}-w)}\mathcal{O}_{\Delta}(\chi,\rho)=\tilde{C}_{\mathcal{O}}\int_{-\infty}^{\infty}\int_{-\infty}^{\infty}d\chi\, d\rho\, e^{+i(\omega_\chi)\,\chi}\mathcal{O}_{\Delta}(\chi,\rho)\,,
\end{equation}
where we have identified the Fourier mode $\omega_{\chi}=-i\Delta_{TB}$ corresponding to $\chi$. Using the commutator \eqref{eq:tzcomm}, one can once again explicitly check that $\omega_{\chi}=-i\Delta_{TB}$ is precisely the eigenmode of the modular Hamiltonian $H_{mod}$ 
\begin{equation}\label{eq:HmodBcomm}
	\left[H_{mod},\mathcal{B}_{\mathcal{O}\mathcal{O}\mathcal{O}_\Delta}(T,B)\right]=-2\pi\,\omega_\chi\,\mathcal{B}_{\mathcal{O}\mathcal{O}\mathcal{O}_\Delta}(T,B)=2\pi\,i\, \Delta_{TB}\,\mathcal{B}_{\mathcal{O}\mathcal{O}\mathcal{O}_\Delta}(T,B).
\end{equation}
Therefore, we can finally write down \eqref{eq:opeeqk4} as
\begin{equation}\label{eq:opeeqk3}
	\mathcal{B}_{\mathcal{O}\mathcal{O}\mathcal{O}_\Delta}(T,B)=\tilde{C}_{\mathcal{O}}\int_{-\infty}^{\infty}\int_{-\infty}^{\infty}d\chi\, d\rho\, e^{+i\omega_\chi\,\chi}\mathcal{O}_{\Delta}(\chi,\rho)=4\pi^2 \tilde{C}_{\mathcal{O}}\,\mathcal{O}_{\Delta}(\omega_{\chi}=-i\Delta_{TB},k_{\rho}=0)\,.
\end{equation}
Note the difference between \eqref{eq:HBcomm} and \eqref{eq:HmodBcomm}. 

For completeness and future use, let us also point out that 
\begin{equation}\label{eq:HmodBcomm2}
	\left[H_{mod},\mathcal{B}_{\mathcal{O}_{\ell_L}\mathcal{O}_{\ell_R}\mathcal{O}_\Delta}(L,R)\right]=2\pi\,i\, \ell_{LR}\,\mathcal{B}_{\mathcal{O}_{\ell_L}\mathcal{O}_{\ell_R}\mathcal{O}_\Delta}(L,R)\,,
\end{equation}
in case there is a spin-difference between the left and right external operators. Similarly
\begin{equation}\label{eq:HBcomm2}
	\left[P_{mod},\mathcal{B}_{\mathcal{O}_{\ell_B}\mathcal{O}_{\ell_T}\mathcal{O}_\Delta}(T,B)\right]=2\pi\,i\, \ell_{BT}\,\mathcal{B}_{\mathcal{O}_{\ell_B}\mathcal{O}_{\ell_T}\mathcal{O}_\Delta}(T,B)\,,
\end{equation}
if there are spin differences between the top-bottom external operators. Similar expressions appear in appendix \ref{app:HBcomm}, especially when the external operators are stress-tensor and scalar respectively. See also \cite{Das:2019iit} where similar commutators were considered for the left-right OPE blocks.

\subsection{Scalar channel OPE blocks between scalar and stress tensor}\label{subsec:phiTope}

We will now similarly study the OPE blocks in the left-right and top-bottom case, with one of the external operators having non-zero spin. Even though our results can be easily generalized (at least for CFT$_2$) to higher spin conserved currents, for now, we will only consider stress tensor to be the spinning external operator. Our convention will be to place the spinning operator either on the left, or on the bottom endpoint of the CD.

\noindent
\emph{Left-right blocks}:
In the first case, using the shadow operator prescription, the OPE block takes the form 
\begin{eqnarray*}
	\mathcal{B}_{T_{\mu\nu}\mathcal{O}\mathcal{O}_\Delta}(L,R)
	=C_{\mathcal{O}}\int_{D_{z\zb}}dz\,d\zb\,\left\langle T_{\mu\nu}(L,\bar{L})\mathcal{O}_{\Delta_R}(R,\bar{R})\tilde{\mathcal{O}}_{\tilde{\Delta}}(z,\zb)\right\rangle\mathcal{O}_{\Delta}(z,\zb)
\end{eqnarray*}
\begin{align}\label{eq:opebtoolr}
	=\tilde{C}_{\mathcal{O}}\int_{D_{z\zb}}dz\,d\zb\, \left(\frac{(z-L)(\zb-L)}{(R-z)(R-\zb)}\right)^{h_R-h_L+\ell_{LR}}&\left(\frac{(z-L)(R-z)}{R-L}\right)^{h-1}\left(\frac{(\zb-R)(L-\zb)}{R-L}\right)^{\bar{h}-1}\nonumber\\
	&\times\left(Z_{\mu}Z_{\nu}-\text{traces}\right)\,\mathcal{O}_{\Delta}(z,\zb)\,.
	\nonumber\\
\end{align}
Here the $\left(Z_{\mu}Z_{\nu}-\text{traces}\right)$ indicates the traceless piece that appears in a three-point function of type $\langle T\mathcal{O}\mathcal{O}\rangle$ (see e.g. \cite{Simmons-Duffin:2016gjk}) 
\begin{equation}\label{eq:Zmuexprssn0}
	Z^\mu(z,\zb)=\frac{x_{Lx}^\mu}{x_{Lx}^2}-\frac{x_{LR}^\mu}{x_{LR}^2},
\end{equation}
where $x^\mu_{ij}=x^\mu_i-x^\mu_j$ and the $x$ in the subscript indicates a generic point within the causal diamond.\footnote{For example, if the higher spin current is a stress tensor, then the traceless factor in the numerator is 
	\begin{equation}\label{eq:tracelesspartcft2}
		\left(Z_{\mu_1}\dots Z_{\mu_{\ell_B}}-\text{traces}\right)\to Z_\mu Z_\nu-\frac{1}{d}\,g_{\mu\nu}Z^2
	\end{equation}
	\label{foot:TLHS}
	in CFT$_d$ \cite{Osborn:1993cr}.} Plugging everything in lightcone coordinates and converting other terms to Rindler, we obtain (here we have only given the expression for OPE blocks involving $T_{\zb\zb}$, but similar expressions can also be obtained for $zz$ component)
\begin{align}\label{eq:opetoolrrind}
	\mathcal{B}_{T_{\zb\zb}\mathcal{O}\mathcal{O}_\Delta}(L,R)
	&=\tilde{C}_{\mathcal{O}}\int_{-\infty}^{\infty}\int_{-\infty}^{\infty}d\chi\, d\rho\, e^{-i(-i\Delta_{LR})\,\rho}\, e^{i(i\ell_{LR})\,\chi}\,\mathcal{O}_{\Delta}(\chi,\rho)\nonumber\\
	&=\tilde{C}_{\mathcal{O}}\int_{-\infty}^{\infty}\int_{-\infty}^{\infty}d\chi\, d\rho\, e^{-ik_{\rho}\,\rho}\, e^{i\omega_\chi\,\chi}\,\mathcal{O}_{\Delta}(\chi,\rho)\,.
\end{align}
We can then interpret the above OPE block as the following mode of the boundary field
\begin{equation}\label{eq:opeeqk7}
	\mathcal{B}_{T_{\zb\zb}\mathcal{O}\mathcal{O}_\Delta}(L,R)=4\pi^2 \tilde{C}_{\mathcal{O}}\,\mathcal{O}_{\Delta}(\omega_{\chi}=i\ell_{LR},k_{\rho}=-i\Delta_{LR})\,.
\end{equation}
We see that the above relations are consistent with \eqref{eq:HBcomm} and \eqref{eq:HmodBcomm2}, which is also obtained via direct computation in appendix \ref{app:HBcomm}. 

\noindent
\emph{Top-bottom blocks}:
In a similar manner, we can also study the case of $\mathcal{B}_{T_{\mu\nu}\mathcal{O}\mathcal{O}_\Delta}(T,B)$. In $(z,\zb)$ plane it takes the form
\begin{eqnarray*}
	\mathcal{B}_{T_{\mu\nu}\mathcal{O}\mathcal{O}_\Delta}(T,B)
	=C_{\mathcal{O}}\int_{D_{z\zb}}dz\,d\zb\,\left\langle T_{\mu\nu}(B,\bar{B})\mathcal{O}_{\Delta_T}(T,\bar{T})\tilde{\mathcal{O}}_{\tilde{\Delta}}(z,\zb)\right\rangle\mathcal{O}_{\Delta}(z,\zb)
\end{eqnarray*}
\begin{align}\label{eq:opebtootb}
	=\tilde{C}_{\mathcal{O}}\int_{D_{z\zb}}dz\,d\zb\, \left(\frac{(B-z)(\zb-\bb)}{(z-T)(\tb-\zb)}\right)^{\frac{\ell_{BT}+\Delta_T-\Delta_B}{2}}
	&\left(\frac{(z-T)(B-z)}{B-T}\right)^{h-1}\left(\frac{(\zb-\bar{B})(\bar{T}-\zb)}{\bar{T}-\bar{B}}\right)^{\bar{h}-1}\nonumber\\
	&\times\left(Z_{\mu}Z_{\nu}-\text{traces}\right)\,\mathcal{O}_{\Delta}(z,\zb)\,.\nonumber\\
\end{align}
Once again, converting these equations in Rindler we have (once again, here we have only given the expression for OPE blocks involving $T_{\zb\zb}$, but similar expressions can also be obtained for $zz$ component)
\begin{align}\label{eq:opeeqk5}
	\mathcal{B}_{T_{\zb\zb}\mathcal{O}\mathcal{O}_\Delta}(T,B)&=\tilde{C}_{\mathcal{O}}\int_{-\infty}^{\infty}\int_{-\infty}^{\infty}d\chi\, d\rho\, e^{-\frac{\ell_{BT}}{2}(\bar{w}+w)}\,e^{-\frac{\Delta_{BT}}{2}(\bar{w}-w)}\mathcal{O}_{\Delta}(\chi,\rho)\nonumber\\
	&=\tilde{C}_{\mathcal{O}}\int_{-\infty}^{\infty}\int_{-\infty}^{\infty}d\chi\, d\rho\,e^{-ik_{\rho}\,\rho}\, e^{i\omega_{\chi}\,\chi}\mathcal{O}_{\Delta}(\chi,\rho)\,.
\end{align}
As before, they are consistent with \eqref{eq:HmodBcomm} and \eqref{eq:HBcomm2}, as we have also shown via direct computation in appendix \ref{app:HBcomm}. Therefore, we can interpret the above OPE blocks as the following mode of the boundary field
\begin{equation}\label{eq:opeeqk6}
	\mathcal{B}_{T_{\zb\zb}\mathcal{O}\mathcal{O}_\Delta}(T,B)=4\pi^2 \tilde{C}_{\mathcal{O}}\,\mathcal{O}_{\Delta}(\omega_{\chi}=-i\Delta_{TB},k_{\rho}=-i\ell_{BT})\,.
\end{equation}
We note that in both \eqref{eq:opeeqk7} and \eqref{eq:opeeqk6}, the modes are Fourier modes only if $\ell_{LR}, \ell_{BT}, \Delta_{LR}$ and  $\Delta_{TB}$ are all purely imaginary. However, as we will see in the next section, if we want to interpret such OPE blocks as geodesic bulk fields, then we must have $\ell_{LR}=\Delta_{TB}=0$. This is expected, as the bulk fields located on the RT surface are always zero modes of the modular Hamiltonian.

\section{Reconstructing bulk with OPE blocks}\label{sec:opehkllmatch}

Combining the results of sections \ref{sec:RHKLL} and \ref{sec:ope}, we now see that in certain cases, the CFT$_2$ OPE blocks naturally furnish a representation for the geodesic bulk operators. For the Rindler scalars, the HKLL prescription leads to a bulk-boundary map given by   \eqref{eq:fttransformedmap-new}, which we have rewritten here for reader's convenience:
\begin{equation}\label{eq:fttransformedmap2}
	\mathcal{O}_{\Delta}(k_{\rho})=\left(\frac{2^{\Delta-1}(\Delta-1)|\Gamma\left(\frac{\Delta}{2}+\frac{ik_\rho}{2}\right)|^2}{4\pi^2C_{\Delta}\Gamma(\Delta)}\right)\,\int_{-\infty}^{\infty}d\rho\,e^{-ik_{\rho}\rho}\,\phi_{Rindler}(\tilde{r}=1,\rho)\,.
\end{equation}
Comparing it with the OPE blocks \eqref{eq:opeeqk2}, we indeed find the well-known relation \cite{daCunha:2016crm,Czech:2016xec,deBoer:2016pqk}
\begin{align}\label{eq:opebbulkrel-new1}
	\frac{1}{4\pi^2\tilde{C}_{\mathcal{O}}}\mathcal{B}_{\mathcal{O}\mathcal{O}\mathcal{O}_\Delta}(L,R)&=\mathcal{O}_{\Delta}(\omega_{\chi}=0,k_{\rho}=-i\Delta_{LR})\nonumber\\
	&=\left(\frac{2^{\Delta-1}(\Delta-1)|\Gamma\left(\frac{\Delta}{2}+\frac{\Delta_{LR}}{2}\right)|^2}{4\pi^2C_{\Delta}\Gamma(\Delta)}\right)\int_{-\infty}^{\infty}d\rho\,e^{-ik_{\rho}\rho}\big{|}_{k_{\rho}=-i\Delta_{LR}}\,\phi_{Rindler}(\tilde{r}=1,\rho)\,.
\end{align}
In other words, in order to identify the above OPE block with the Fourier modes of the geodesic bulk field, we need the imaginary $k_{\rho}$ modes of the boundary primary.

However, similar relations between the OPE blocks and Rindler scalars continue to exist even for other OPE blocks (left-right or top-bottom) given by \eqref{eq:opeeqk3}, \eqref{eq:opeeqk7} and \eqref{eq:opeeqk6} as long as the modular energy modes are taken to zero (i.e. with $\Delta_{TB}=0$ in \eqref{eq:opeeqk3} and \eqref{eq:opeeqk6}, and $\ell_{LR}=0$ in \eqref{eq:opeeqk7}). In particular, whenever the OPE block is given by (where depending on the type of the OPE block, $f,g$ take values such as $i\Delta_{LR}$, $i\Delta_{TB}$, $i\ell_{LR}$ or $i\ell_{BT}$)
\begin{equation}
	\mathcal{B}=4\pi^2\,\tilde{C}_{\mathcal{O}}\, \mathcal{O}_\Delta(\omega_\chi=f,k_\rho=g) \qquad\text{with}\qquad f=0\,,
\end{equation}
we have
\begin{equation}
	\int_{-\infty}^{\infty}d\rho\,e^{-ik_{\rho}\rho}\,\phi_{Rindler}(\tilde{r}=1,\rho)\propto \mathcal{B}\,.
\end{equation}
Because modular Hamiltonians always pick out the modular energy values as the eigenvalues for the boundary operators, we see that in all cases, the geodesic bulk fields are modular zero modes. We also notice that for $g=i\Delta_{LR}$ or $i\Delta_{TB}$, it is a Fourier mode only for external operators taken from the principal series representation of the CFT. This is naturally obtained in the context of Euclidean CFTs relevant for dS/CFT. This leads us to investigate whether a similar dictionary exists between the dS bulk fields and Euclidean OPE blocks or not. This will be the topic of our next section. Note that in the above cases, one might even be tempted to suggest that the dictionary between the OPE blocks and the bulk fields are more fundamental for dS/CFT, and the case for AdS/CFT is understood as an analytic continuation from dS. During this analytic continuation from dS to AdS, the external operators remain from the principal series, although the primary operator in the channel switches to real conformal dimension. It will be interesting to see if such an interpretation is justified.

\section{de Sitter holography using OPE blocks}\label{sec:dSconn}

In this section, we elaborate upon the connection between bulk scalar fields in dS flat patch, and the role of Euclidean OPE blocks (reconstruction of higher spin fields go in a similar manner, see \cite{Sarkar:2014dma}). These OPE blocks naturally incorporate operators from continuous series representation depending on the mass-squared value of the bulk fields in question \cite{Chatterjee:2015pha,Chatterjee:2015wva,Chatterjee:2016ifv}. The flat slicing of dS can be obtained by a simple analytic continuation of the AdS Poincar\'{e} patch \eqref{eq:pads} in the following manner
\be\label{eq:analc}
Z\rightarrow \eta \; ;\, t\rightarrow t \;;~ x \rightarrow ix~~ \text{and~~} R_{AdS} \rightarrow iR_{dS}\;~ \text{with}~\, m^2R_{AdS}^2\rightarrow-m^2R_{dS}^2\,,
\ee
which leads to the de-Sitter static patch metric 
\be \label{eq:dSflatmetric}
ds^2 = \frac{R_{dS}^2}{\eta^2}(-d\eta^2+dt^2+dx^2)\,,
\ee  
Note that $\eta$ is now the new timelike coordinate. In these coordinates, the bulk scalar fields are now written as a smearing of both sources and vevs \cite{Xiao:2014uea}
\begin{align}\label{eq:dsflat}
	\Phi(\eta,x^\mu)&=C \int_{\abs{y}^2<\eta^2} d^2y \left(\frac{\eta^2-\abs{y}^2}{\eta}\right)^{\Delta-2} \mathcal{O}_{+}\left(x^\mu+y^\mu\right)+C' \int_{\abs{y}^2<\eta^2} d^2y\left(\frac{\eta^2-\abs{y}^2}{\eta}\right)^{-\Delta} \mathcal{O}_{-}\left(x^\mu+y^\mu\right)\,, \nonumber\\
\end{align}
where $C$ and $C'$ are constants depending upon the value of $\Delta$ and the space-time dimensions. For convenience we write the above equation in short as
\begin{equation}
	\Phi(\eta,x,t)=\Phi_+(\eta,x,t)+\Phi_-(\eta,x,t).
\end{equation}
The analytic continuation takes the boundary geometry from Lorentzian to that of an Euclidean CFT. One of the features of Euclidean CFTs is that they have a continuous spectrum of operators with complex weights, unlike their Lorentzian counterparts. The continuous series representation \cite{Chatterjee:2015pha} is a way of describing the Hilbert space of such operators in terms of harmonic analysis on the conformal group. Depending on the mass parameter $m^2$, there are two types of continuous series: complementary and principal. The complementary series occurs when $m^2R_{dS}^2<\frac{d^2}{4}$, and the principal series when $m^2R_{dS}^2>\frac{d^2}{4}$.\footnote{The discrete series representation is another possibility for Euclidean CFTs, but it is not relevant for our study.} The conformal weights of these latter operators take the form $\Delta=\frac{d}{2} + i\sqrt{m^2R_{dS}^2-\frac{d^2}{4}}$, and as a result, depending on the mass of the associated bulk field, $\Delta$ can either be complex or real. In the former case they belong to the principal series representation \cite{Gadde:2017sjg,Chatterjee:2015pha}, and the latter is known as the complementary series representation. We shall only stick to principal series represented operators for our purpose.
\begin{figure}[t]
	\begin{center}
		\includegraphics[width=135mm]{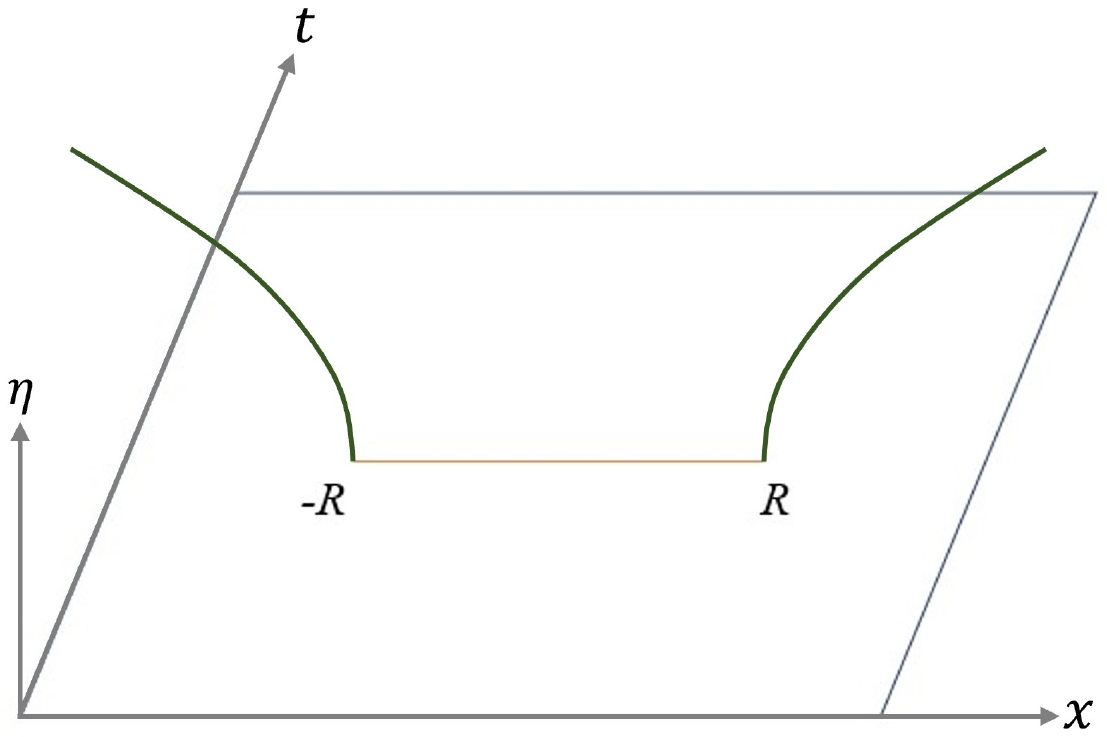}
		\caption{Analogue of RT surface in de Sitter for a boundary subregion $x\in [-R,R]$}
		\label{fig:dSRT}
	\end{center}
\end{figure}
Also notice that the analytic continuation transforms the RT surface equation to $x^2=\eta^2+R^2$, which is indeed the analogous surface in dS space (see figure \ref{fig:dSRT}) that captures the pseudo-entropy \cite{Doi:2022iyj} for a subregion extending from $x=-R$ to $x=R$. However, unlike AdS, this surface does not turn in the bulk; rather they are two disjoint surfaces anchored at $x=\pm R$. Only in the global patch of dS they can be connected \cite{Doi:2022iyj}. In order to show our claim that the boundary Euclidean OPE blocks are bulk fields integrated over these surfaces, we need to work with shadow prescription in the momentum space of the boundary coordinate for the channel operator. This was already used in \cite{Chen:2019fvi}, to derive a \emph{new} representation of smeared higher dimensional bulk fields in terms of Lorentzian OPE blocks in AdS. However, here our goal is to derive a connection between geodesic, local bulk fields in dS with Euclidean OPE blocks.

We begin by expressing the OPE block for external operators at positions \( x_1 \) and \( x_2 \) using the shadow formalism. By analytically continuing our Lorentzian shadow representation \eqref{eq:opeb} to the Euclidean plane,\footnote{For a recent study on Euclidean OPE blocks as a position space integral over CFT coordinates, see \cite{Agon:2024vif}.} the integration domain for the Euclidean case naturally extends to all of space, as the concept of causality is lost upon transitioning to Euclidean signature. Thus, we write the Euclidean OPE block in shadow formalism as
\begin{equation}\label{eq:EOPEdef}
	\mathcal{B}^E_{\mathcal{O}\mathcal{O}\mathcal{O}_{\Delta}}(x_1, x_2) = c_{\Delta} \int_{\text{all space}} d^d x \, \langle \mathcal{O}(x_1) \mathcal{O}(x_2) \tilde{\mathcal{O}}_{\bar{\Delta}}(x) \rangle \, \mathcal{O}_{\Delta}(x)\,.
\end{equation}

However, working directly in position space is not convenient for expressing the result in terms of the holographic degrees of freedom. To address this, we develop the shadow prescription for Euclidean OPE blocks in momentum space (see Appendix~D). Under this prescription, we start by performing a Fourier transform of \eqref{eq:EOPEdef} thereby obtaining the OPE block in momentum space as

\begin{equation}\label{eq:eOPE}
	\mathcal{B}_{\mathcal{O}\mathcal{O}\mathcal{O}_{\Delta,J}}^E(x_1,x_2) = \frac{1}{\alpha_{\Delta,J}\alpha_{\Bar{\Delta},J}} \int d^dp\, E_{\Delta_1,\Delta_2,\Bar{\Delta},J} (x_1,x_2,-p;z_3) E_{{\Bar{\Delta},J}}(-p)\mathcal{O}_{\Delta,J}(p)\,.
\end{equation}
Here $\Delta$ and $J$ are the conformal weight and spin of the channel operator, $\alpha_{\Delta,J}$ and $\alpha_{\Bar{\Delta},J}$ are gamma functions of the arguments, and $z_3$ is the corresponding polarization.\footnote{Here we have used $\Bar{\Delta}=d-\Delta~,~\bar{d}=\frac{d}{2}$ and $\alpha_{\Delta,J}=2^{d-2\Delta}\pi^{h}\frac{(\Delta-1)_J\Gamma(h-\Delta)}{\Gamma(\Delta+J)}$. Also, we'll use the notation $\tau=\Delta-J$ and $\Bar{\tau}=\Bar{\Delta}-J$\,.\;$\Delta_{12}^\pm=\Delta_1\pm\Delta_2$ together with $\delta_{12}^{\pm}=\frac{\tau\pm\Delta_{12}^-}{2}$, $\Bar{\delta}_{12}^{\pm}=\frac{\Bar{\tau}\pm\Delta_{12}^-}{2}$ and $\mathcal{N}_{12,\Delta,J}=\frac{1}{\Gamma(\delta_{12}^\pm+J)}.$\label{foot:alphadef}} Here the Euclidean three-point function is denoted by $E_{\Delta_1,\Delta_2,\Bar{\Delta},J} (x_1,x_2,-p;z_3)$, and $E_{{\Bar{\Delta},J}}(-p)$ is the Euclidean two-point function. In what follows we derive the momentum three-point function by Fourier transforming the coordinates associated with the channel operator. The derivation follows as
\begin{equation}
	E_{\Delta_1,\Delta_2,\Bar{\Delta},J}\left(x_1,x_2,-p;z_3\right)
	= \int d^dx_3\, E_{\Delta_1,\Delta_2,\Bar{\Delta},J}\left(x_1,x_2,x_3;z_3\right)e^{ip\cdot x_3}\label{eq:momentum3pt} 
\end{equation}
\begin{align}
	&=\frac{\pi^{\bar{d}}}{2^{J-1}}\frac{\mathcal{N}_{12,\Delta,J}}{\left(x_{12}^{2}\right)^{\frac{\Delta_{12}^{+}-\tau}{2}}} \mathscr{D}_J \left(\delta_{12}^{+},z_3 \cdot \partial_{x_1} ; \delta_{12}^{-},z_3 \cdot \partial_{x_2}\right)\left(\frac{p^2}{4 {x_{12}^{2}}}\right)^{\frac{\Delta-J-{\bar{d}}}{2}} \int_0^1 du\,u^{\frac{\Delta_{12}^{-}+{\bar{d}}}{2}-1} \left(1-u\right)^{\frac{{\bar{d}}-\Delta_{12}^{-}}{2}-1}  \nonumber\\ 
	& \qquad\qquad\qquad\qquad \qquad\qquad\qquad\qquad\times e^{ip\cdot\left(ux_1+\left(1-u\right)x_2\right)}\, K_{{\bar{d}}-\Delta+J} \left(\sqrt{u\left(1-u\right)p^2x_{12}^2}\right)\,,\label{eq:bessel}
\end{align}
where $K$ is the modified Bessel function of second kind. In going from \eqref{eq:momentum3pt} to \eqref{eq:bessel}, we used Feynman-like parametrization as well as the differential operator $\mathscr{D}_J$ discussed in appendix \ref{sec:OPEdS}. 
Using \eqref{eq:bessel} and defining $x\left(u\right)=ux_1+\left(1-u\right)x_2$ we get
\begin{align}
	&E_{\Delta_1,\Delta_2,\Bar{\Delta},J}\left(x_1,x_2,-p;z_3\right)
	=\frac{\pi^{{\bar{d}}+1}\mathcal{N}_{12,\Bar{\Delta},J}}{2^J\sin\left(\pi\left({\bar{d}}-\tau\right)\right)}\Biggl[\frac{1}{\left(x_{12}^2\right)^{\frac{\Delta_{12}^{+}}{2}-{\frac{\tau}{2}}}} \mathscr{D}_J \left(\delta_{12}^{+},z_3 \cdot \partial_{1} ; \delta_{12}^{-},z_3 \cdot \partial_{2}\right) \left(\frac{p^2}{4{x_{12}^{2}}}\right)^{\frac{\tau-{\bar{d}}}{2}}\nonumber \\
	& \int_0^1 du\,u^{\frac{\Delta_{12}^{-}+{\bar{d}}}{2}-1} \left(1-u\right)^{\frac{{\bar{d}}-\Delta_{12}^{-}}{2}-1} e^{ip \cdot x\left(u\right)}\Bigl(I_{\left(\tau-{\bar{d}}\right)}\left(\sqrt{u\left(1-u\right)p^2x_{12}^2}\right)-I_{\left({\bar{d}}-\tau\right)}\left(\sqrt{u\left(1-u\right)p^2x_{12}^2}\right) \Bigr)\Biggr]\,.
\end{align} 
Now we use the identity \eqref{eq:id1} derived in \cite{Chen:2019fvi} along with \eqref{eq:Qdef} which leads to
\begin{align}
	&E_{\Delta_1,\Delta_2,\Bar{\Delta},J}\left(x_1,x_2,-p;z_3\right)\nonumber\\
	&=\frac{\pi^{{\bar{d}}+1}}{2^J}\frac{\mathcal{N}_{12,\Bar{\Delta},J}}{\sin\left(\pi\left({\bar{d}}-\tau\right)\right)}\Biggl[-\kappa_{\Delta,J}E_{\Delta,J}\left(-p\right)\mathcal{Q}_{\Delta_1,\Delta_2,\Bar{\Delta},J}\left(x_1,x_2,-p\right)+\mathcal{Q}_{\Delta_1,\Delta_2,\Delta,J}\left(x_1,x_2,-p\right)\Biggr]\,.
\end{align}
Using the Fourier transform
\begin{align}
	E_{\Delta_1,\Delta_2,\Delta,J}\left(x_1,x_2;x_3,z_3\right)&=\int_{-\infty}^{+\infty}\frac{d^{d-1}\Vec{p}}{\left(2\pi\right)^{d-1}}e^{i\Vec{p}\cdot\Vec{x}} \int_{-\infty}^{+\infty} \frac{dp^0}{2\pi}e^{ip^0 \cdot x^0} E_{\Delta_1,\Delta_2,\Delta,J}\left(x_1,x_2,p;z_3\right)\,,
\end{align}
we see that the contribution to the integral on the right hand side comes from the  upper-half $p$-plane, where the branch cut is from $i\abs{\vec{p}}$ to $i\infty$. The contribution can be obtained by deforming the contour to wrap around this cut \cite{Bautista:2019qxj}. But the $\mathcal{Q}-$kernel is an entire function of $p$ (as it has a Bessel function polynomial expansion) and hence the contribution to the branch integral only comes from the second term in $E_{\Delta_1,\Delta_2,\Delta,J}\left(x_1,x_2,p;z_3\right)$ which include the 2-pt function $E_{\Delta,J}\left(-p\right)$. Therefore,

\begin{align}
	E_{\Delta_1,\Delta_2,\Delta,J}\left(x_1,x_2,x_3;z_3\right) &=\int \frac{d^dp}{\left(2\pi\right)^d}e^{ip\cdot x} \Biggl[-\frac{\pi^{{\bar{d}}+1}}{2^J}\frac{\kappa_{\Delta,J}\mathcal{N}_{12,\Delta,J}}{\sin\left(\pi\left({\bar{d}}-\tau\right)\right)}\mathcal{Q}_{\Delta_1,\Delta_2,\Bar{\Delta},J}\left(x_1,x_2,p\right)E_{\Delta,J}\left(p\right)\Biggr]  \nonumber \\
	&=-\frac{\pi^{{\bar{d}}+1}}{2^J}\frac{\kappa_{\Delta,J}\mathcal{N}_{12,\Delta,J}}{\sin\left(\pi\left({\bar{d}}-\tau\right)\right)}\mathcal{Q}_{\Delta_1,\Delta_2,\Bar{\Delta},J}\left(x_1,x_2,p\right)E_{\Delta,J}\left(p\right)\,.
\end{align}
Thus (here $x_{12}=|x_1-x_2|$),
\begin{align}
	\mathcal{B}_{\mathcal{O}\mathcal{O}\mathcal{O}_{\Delta,J}}^E\left(x_1,x_2\right) =&\frac{\pi^{{\bar{d}}+1}}{2^J}\frac{\kappa_{\Delta,J}\mathcal{N}_{12,\Delta,J}}{\sin\left(\pi\left({\bar{d}}-\Delta\right)\right)}\int d^dp \, \frac{1}{\left(x_{12}^2\right)^{\frac{\Delta_{12}^{+}}{2}-\Bar{\Delta}}}\left(\frac{-p^2}{4{x_{12}^{2}}}\right)^{\frac{\Bar{\Delta}-{\bar{d}}}{2}}\int_0^1 du\,u^{\frac{\Delta_{12}^{-}+{\bar{d}}}{2}-1} \nonumber \\
	&\qquad \qquad\times\left(1-u\right)^{\frac{{\bar{d}}-\Delta_{12}^{-}}{2}-1} e^{ip \cdot x\left(u\right)}J_{\bar{d}-\Bar{\Delta}}\left(i\sqrt{u\left(1-u\right)p^2x_{12}^2}\right)\mathcal{O}_{\Delta,J}(p)\nonumber\\
	=&\frac{\pi^{{\bar{d}}+1}}{2^{{\bar{d}}-\Delta}}\frac{\kappa_{\Delta,J}\mathcal{N}_{12,\Delta,J}}{\sin\left(\pi\left({\bar{d}}-\Delta\right)\right)}\int_0^1 du\,u^{\frac{\Delta_{12}^{-}}{2}-1} \left(1-u\right)^{-\frac{\Delta_{12}^{-}}{2}-1}\frac{\left(-1\right)^{\frac{\Delta}{2}}}{\left(x_{12}^2\right)^{\frac{\Delta_{12}^{+}}{2}}} \int d^dp\, e^{ip\cdot x\left(u\right)}\nonumber \\
	&\qquad\quad\times \eta(u)^{\bar{d}}\left(\sqrt{p^2}\right)^{{\bar{d}}-\Delta}J_{\Delta-{\bar{d}}}\left(\sqrt{p^2}\,\eta\left(u\right)\right)\mathcal{O}_{\Delta,J}(p)\,.
\end{align}
In the above, we've identified $\eta\left(u\right)=\sqrt{-u\left(1-u\right)x_{12}^2}$. The Bessel functions have the integral representation as follows
\begin{equation}
	J_{\nu}\left(\abs{p}x\right)=\frac{1}{2^\nu \pi^{\bar{d}} \Gamma\left(\nu-\bar{d}+1\right)} \Biggl(\frac{|p|}{x}\Biggr)^{\nu} \int_{|y|\le x} d^dy \left(x^2-|y|^2\right)^{\nu-{\bar{d}}} e^{ip\cdot y}\,,
\end{equation}
with $\nu=\Delta-\bar{d}$. We want to focus on the simple case for $d=2,\,J=0$,
where we have
\begin{align}\label{eq:54}
	\mathcal{B}_{\mathcal{O}\mathcal{O}\mathcal{O}_{\Delta}}^E\left(x_1,x_2\right)
	=\tilde{C}_\Delta\int_0^1 du\,u^{\frac{\Delta_{12}^{-}}{2}-1} \left(1-u\right)^{\frac{-\Delta_{12}^{-}}{2}-1}\int_{|y|\le\eta\left(u\right)} d^2y\Biggl(\frac{\eta^2\left(u\right)-|y|^2}{\eta\left(u\right)}\Biggr)^{\Delta-2}\mathcal{O}_{\Delta}\left(x^\mu\left(u\right)+y^\mu(u)\right)	\,.
\end{align}
Here, $\tilde{C}_{\Delta}=\frac{\pi \kappa_{\Delta,0}\mathcal{N}_{12,\Delta}}{\sin\left(\pi\left(\bar{d}-\Delta\right)\right)\Gamma\left(1-\Bar{\Delta}\right)}\left(\frac{1}{x_{12}^2}\right)^{\frac{\Delta_{12}^{+}}{2}}$. In above, we used the momentum integral to invert the Fourier transform $\mathcal{O}\left(p\right)$. Notice that the second integral actually matches completely to the positive frequency part of the dS scalar field \eqref{eq:dsflat}. Identifying the holographic coordinates and using a slightly different variable $u$ given by
\begin{equation}
	x^\mu\left(u\right)=ux_1^\mu+\left(1-u\right)x_2^\mu\,; \qquad\text{and}\qquad u=\frac{1}{1+e^{2\lambda}}\,,
\end{equation}
we finally arrive at
\begin{equation}\label{eq:dSpositiveF}
	\frac{1}{2\tilde{C}_E}\mathcal{B}_{\mathcal{O}\mathcal{O}\mathcal{O}_{\Delta,J}}^E\left(x_1,x_2\right)
	=\int_{-\infty}^{\infty}\,d\lambda\, e^{\lambda\Delta_{12}^{-}}\,\Phi_{+}\left(\eta\left(u\right),x^\mu\left(u\right)\right)\,,
\end{equation}
where $\tilde{C}_E=\tilde{C}_{\Delta}/C$. The above expression explicitly shows that the boundary Euclidean OPE block in scalar channel is indeed the positive frequency part of the dS bulk scalar field in flat slicing integrated over a surface. To see that this is indeed the analogous RT surface for dS flat slicing we can compute the relation for the holographic coordinates $x^\mu\left(u\right)$ and the bulk radial coordinate $\eta\left(u\right)$, which is given by
\begin{equation}
	x^0\left(u\right)=t\left(u\right)= ux_1^0+\left(1-u\right)x_2^0 ~~~~\text{and}\qquad x^1\left(u\right)=x\left(u\right)=ux_1^1+\left(1-u\right)x_2^1\,.
\end{equation}
Taking the $t=0$ plane, and the boundary subregion to be of length $2R$ stretching from $x_1^1=x(u)=-R$ to $x_2^1=x(u)=R$, we obtain
\begin{align}\label{eq:hypextsurface}
	x^2\left(u\right)-\eta^2\left(u\right)&=\left(ux_1^1+\left(1-u\right)x_2^1\right)^2+u\left(1-u\right)x_{12}^2\nonumber\\
	&=\left(-uR+\left(1-u\right)R\right)^2+4u\left(1-u\right)R^2=R^2\,.
\end{align}
Thus the Euclidean OPE block corresponds to the positive frequency part of the dS bulk field integrated over a hyperbolic surface given by $x^2=\eta^2+R^2$. Indeed we can also write an expression as sum of two OPE blocks corresponding to $\Phi_{+}$ and $\Phi_{-}$ respectively to get the full dS scalar field given by
\begin{equation}
	\frac{1}{2\tilde{C}_E}\mathcal{B}_{\mathcal{O}\mathcal{O}\mathcal{O}_{\Delta}}^{E}(x_1,x_2)+\frac{1}{2\tilde{C}_E'}\mathcal{B}_{\mathcal{O}\mathcal{O}\mathcal{O}_{\Delta}}^{E}(x_1,x_2)
	=\int_{-\infty}^{\infty}\,d\lambda\, e^{\lambda\Delta_{12}^{-}}\,\Phi\,\left(\eta(u),x^\mu(u)\right).
\end{equation}
We note that the OPE blocks are now purely written in terms of operators from the continuous series. So, considering the dS reconstruction to be our starting point, we can now analytically continue back to AdS to recover the HKLL prescription for the bulk field. Note that we have to by-hand impose the vanishing of the non-normalizable part, to ensure the AdS bulk field is normalizable. The backwards  analytic continuation that takes from dS to AdS is given by
\begin{equation} \label{eq:analdSrind}
	x\rightarrow -i x\,,\,t\rightarrow t\,~~ \text{and}~~\eta \rightarrow Z ~~\text{with}~~\, R_{dS}\rightarrow -i R_{AdS}\,,~ m^2R_{dS}^2\rightarrow -m^2R_{AdS}^2\,.
\end{equation}
We notice that the external operators remain from the Principal series representation (which is not inconsistent, as the operator spectrum of $SO(d,2)$ contains the principal series represented operators \cite{Kravchuk:2018htv,Agarwal:2021vqz,Agarwal:2023xwl}), while the channel operator switches from the complex conformal dimension to real, physical representation (which is indeed required for the dual AdS reconstruction). Note that the Principal series represented external operators have conformal dimensions in the form $\Delta=\frac{d}{2}+i\rho$ (for some real value $\rho$), and hence their difference is a purely imaginary quantity. This is consistent with our discussions at the end of section \ref{sec:opehkllmatch}.

\section{Higher dimensional generalizations}\label{sec:higherd}

This section will be devoted to extending our results for the various OPE blocks to higher dimensional AdS/CFT. The generalization is quite straightforward, but from the observations of appendix \ref{subapp:HBcomm}, we will see that the connection between the OPE blocks and the bulk fields on the RT surface aren't anymore connected in a straightforward manner for arbitrary $d$. The main reason behind this failure is the fact that the OPE blocks no longer remain an eigenmode of modular Hamiltonians, unlike the bulk scalars on the RT surface.

\subsection{OPE blocks in higher dimensions}\label{sec:higherdhkll}

In this subsection, we will focus on the OPE block construction in higher dimensions. Similar studies have also appeared previously in \cite{Czech:2016xec,Chen:2019fvi}. Our results are consistent with theirs with some detailed calculations and observations. There are several important differences in higher dimensions. The RT surface $\gamma_S$ is now a codimension-2 hyperbolic surface $H^{d-1}=R^1\ltimes S^{d-2}$, and instead of left ($L$) and right ($R$) endpoints of the boundary causal diamonds, we have  an $S^{d-2}$ surface. In particular, instead of the $(z,\zb)$ plane, we are using the following boundary coordinates 
\begin{equation}\label{eq:higherDp}
	ds^2|_{boundary}=-dt^2+dr^2+r^2d\Omega_{d-2}^2=g_{\mu\nu}\,dx^\mu\, dx^\nu.
\end{equation}
As our entangling region, we consider a spherical subregion of size $r=R$, and we map its associated causal diamond to \cite{Casini:2011kv} 
\begin{equation}\label{eq:higherDr}
	ds^2|_{boundary}=\frac{R^2}{(\cosh\rho+\cosh\chi)^2} (-d\chi^2+d\rho^2+\sinh^2\rho\, d\Omega_{d-2}^2)=\Omega^2\, g^R_{\mu\nu}\,dx_R^\mu\, dx_R^\nu\,,
\end{equation}
via the coordinate transformations
\begin{equation}\label{eq:cthigherd}
	r=R\frac{\sinh\rho}{(\cosh\rho+\cosh\chi)},\, t=R\frac{\sinh\chi}{(\cosh\rho+\cosh\chi)},\,\Omega_{d-2}=\Omega_{d-2}\quad\text{with}\quad \Omega=\frac{R}{(\cosh\rho+\cosh\chi)}.
\end{equation}
In order to explore OPE blocks in such higher dimensional settings, one way to proceed is to place some surface operators along the entangling surface of the spherical subregion \cite{Czech:2016xec,deBoer:2016pqk}, but as we want local CFT primary operators to be our input, this is not a suitable approach. A better choice are the top ($T$) and bottom ($B$) points of the aforementioned causal diamond, along the lines considered in section \ref{sec:ope}. In fact, in this case, we will only study the OPE blocks coming from having a spinning primary $J_{\mu_1\dots\mu_{\ell_B}}$ at point $B$ (we will only consider a stress-tensor for simplicity) and a scalar primary located at $T$. We will denote their conformal dimensions by $\Delta_B$ and $\Delta_T$ respectively, with spins $\ell_B$ and  $\ell_T=0$.\\

The shadow operator formalism in this case gives us the following OPE block expression
\begin{align}\label{eq:opebanyD-2}
	\mathcal{B}_{J\mathcal{O}\mathcal{O}}&=C_d\int_{D}\sqrt{-g}\,dt\,dr\,d\Omega_{d-2}\,\left\langle J^{\mu_1\dots\mu_{\ell_B}}_{\Delta_B,\ell_B}(B)\mathcal{O}_{\Delta_T,\ell_T=0}(T)\tilde{\mathcal{O}}_{\tilde{\Delta},\ell=0}(x)\right\rangle\mathcal{O}_{\Delta,\ell=0}(x)\nonumber\\
	&=\int_{D}\sqrt{-g}\,dt\,dr\,d\Omega_{d-2}\,\frac{C_d\,C_{J\mathcal{O}\tilde{\mathcal{O}}}\,\left(Z^{\mu_1}\dots Z^{\mu_{\ell_B}}-\text{traces}\right)\,\mathcal{O}_{\Delta,\ell=0}(X)\,}{|(x-B)|^{\Delta_B+\tilde{\Delta}-\Delta_T-\ell_B}|(x-T)|^{\Delta_T+\tilde{\Delta}-\Delta_B+\ell_B}|(B-T)|^{\Delta_B+\Delta_T-\tilde{\Delta}-\ell_B}}\,.
\end{align}
Here in the last line we have used the CFT$_d$  three-point function \cite{Simmons-Duffin:2016gjk} and $C_{J\mathcal{O}\tilde{\mathcal{O}}}$ is the corresponding OPE coefficient which is a function of the $\langle\mathcal{O}_{\Delta_T}\tilde{\mathcal{O}}_{\tilde{\Delta}}\rangle$ 2-point function due to conformal Ward identity. Because the scalar two-point function vanishes for different conformal dimensions, this constrains $\Delta_T=\tilde{\Delta}=d-\Delta$. On the other hand, similar to \eqref{eq:Zmuexprssn0}, we have 
\begin{equation}\label{eq:Zmuexprssn}
	Z^\mu=\frac{x_{Bx}^\mu}{x_{Bx}^2}-\frac{x_{BT}^\mu}{x_{BT}^2},
\end{equation}
where $x^\mu_{ij}=x^\mu_i-x^\mu_j$.  
Using $\tilde{\Delta}=d-\Delta$ above, we end up with (below $\tilde{C}_d=C_d\,C_{J\mathcal{O}\tilde{\mathcal{O}}}$)
\begin{equation}\label{eq:opebanyD-2-simp}
	\mathcal{B}_{J\mathcal{O}\mathcal{O}}=\tilde{C}_d\,\int_{D}\sqrt{-g}\,dt\,dr\,d\Omega_{d-2}\,\left(Z_{\mu_1}\dots Z_{\mu_{\ell_B}}-\text{traces}\right)\, I_1\,I_2\,\mathcal{O}_{\Delta,\ell=0}(x),
\end{equation}
where 
\begin{equation}\label{eq:I1}
	I_1=\left(\frac{|x-T||B-x|}{|T-B|}\right)^{\Delta-d} \qquad\text{and}\qquad I_2=\left(\frac{|x-B|}{|x-T|}\right)^{\ell_B+\Delta_T-\Delta_B}\,.
\end{equation}
Here, we've also absorbed a $|B-T|^{-\Delta_T-\Delta_B+\ell_B}$ factor in $\tilde{C}_{d}$. In our notation $x=(t,r,\vec{\phi})$ with $\vec{\phi}$ being the directions along $S^{d-2}$. The symmetry of the situation also guarantees that the top and bottom points are at the same $|\vec{\phi}|$ value, but it happens to be at the origin or pole of the $\vec\phi$ coordinate. This drastically simplifies a lot of expressions, as we do not need to consider the $\vec\phi$ directional distance between $x$ and $(T,B)$.

It is useful to again define and work with lightcone coordinates consisting of the radial direction $r$ and time $t$ \cite{Casini:2011kv}. So we coordinate transform to Poincar\'{e} and Rindler lightcone variables $z$ and $w$ respectively in a manner similar to the AdS$_3$/CFT$_2$ case. In other words, we define 
\begin{equation}\label{eq:higherdzzbardef}
	(z,\zb)=(r-t,r+t)\qquad\text{and}\qquad (w,\wb)=(\rho-\chi,\rho+\chi)\,.
\end{equation}
In these coordinates, we finally have 
\begin{align}\label{eq:TBOPE1}
	\mathcal{B}_{J\mathcal{O}\mathcal{O}}&=\tilde{C}_d\,\int_{D}\sqrt{-g(w)}\,dw\,d\wb\,e^{\chi(\ell_{BT}+\Delta_{TB})}\left[\int\,d\Omega_{d-2}\,\left(Z^{\mu_1}\dots Z^{\mu_{\ell_B}}-\text{traces}\right)\right]\,\mathcal{O}_{\Delta,\ell=0}(w,\wb,\phi)\,\nonumber\\
	&=\tilde{C}_d\,\int_{D}\sqrt{-g^R}\,d\rho\,d\chi\,e^{\chi(\ell_{BT}+\Delta_{TB})}\left[\int\,d\Omega_{d-2}\,\left(Z^{\mu_1}\dots Z^{\mu_{\ell_{B}}}-\text{traces}\right)\right]\,\mathcal{O}_{\Delta,\ell=0}(\rho,\chi,\phi)\,.
\end{align}
In the last line above, just like for the CFT$_2$ case, we have written everything in $\rho,\chi$ coordinates with $g_{\mu\nu}^R$ defined in \eqref{eq:higherDr}. 
If we use the null contracted higher spin currents (see e.g.~section 3.5 of \cite{Kirilin:2018qpy}), then the OPE block expression is
\begin{align}\label{eq:TBOPE2}
	\mathcal{B}_{J\mathcal{O}\mathcal{O}}=\tilde{C}_d\,\int_{D}\sqrt{-g^R}\,d\rho\,d\chi\,e^{\chi(\ell_{BT}+\Delta_{TB})}\left[\int\,d\Omega_{d-2}\,\left(n_\mu\, Z^{\mu}\right)^{\ell_{BT}}\right]\,\mathcal{O}_{\Delta,\ell=0}(\rho,\chi,\phi)\,.
\end{align}
As a quick check of our above expressions, we note that if we are in two-dimensions, and the conserved current is the stress tensor, we precisely recover \eqref{eq:opeeqk5} that we obtained before.

In order to write down explicit expressions, we first resort to taking the higher spin current to be a stress tensor (we will conjecture a form for higher spin, higher dimensional OPE block at the end of this section). The resulting expressions of the OPE block becomes
\begin{equation}\label{btoohighd}
	\mathcal{B}_{T_{zz}\mathcal{O}\mathcal{O}}=\tilde{C}_d\,\int_{D}\sqrt{-g^R}\,d\rho\,d\chi\,e^{\rho\,\ell_{BT}}\,e^{\chi\,\Delta_{TB}}\left[\int\,d\Omega_{d-2}\right]\,\mathcal{O}_{\Delta,\ell=0}(\rho,\chi,\phi)\,,
\end{equation}
and
\begin{equation}\label{btbaroohighd}
	\mathcal{B}_{T_{\zb\zb}\mathcal{O}\mathcal{O}}=\tilde{C}_d\,\int_{D}\sqrt{-g^R}\,d\rho\,d\chi\,e^{-\rho\,\ell_{BT}}\,e^{\chi\,\Delta_{TB}}\left[\int\,d\Omega_{d-2}\right]\,\mathcal{O}_{\Delta,\ell=0}(\rho,\chi,\phi)\,.
\end{equation}
On the other hand computing the spin part of $\langle T_{z\zb}\mathcal{O}\mathcal{O}\rangle$, $\langle T_{z\phi}\mathcal{O}\mathcal{O}\rangle$, $\langle T_{\zb\phi}\mathcal{O}\mathcal{O}\rangle$ and $\langle T_{\phi\phi}\mathcal{O}\mathcal{O}\rangle$
we have
\begin{equation}\label{bttbaroohighd}
	\mathcal{B}_{T_{z\zb}\mathcal{O}\mathcal{O}}=\tilde{C}_d\left[\frac{1}{4R^2}\left(\frac{1}{4}-\frac{1}{2d}\right)\right]\,\int_{D}\sqrt{-g^R}\,d\rho\,d\chi\,e^{\chi\,\Delta_{TB}}\left[\int\,d\Omega_{d-2}\right]\,\mathcal{O}_{\Delta,\ell=0}(\rho,\chi,\phi)\,,
\end{equation}
which clearly vanishes for $d=2$. On the other hand,
\begin{equation}\label{btppoohighd}
	\mathcal{B}_{T_{\phi\phi}\mathcal{O}\mathcal{O}}=-\frac{\tilde{C}_d}{4dR^2}\,\int_{D}\sqrt{-g^R}\,d\rho\,d\chi\,e^{\chi\,\Delta_{TB}}\,\left(\frac{z+\zb}{2}\right)^2\,\left[\int\,d\Omega_{d-2}\right]\,\mathcal{O}_{\Delta,\ell=0}(\rho,\chi,\phi)\,,
\end{equation}
where the factor $\left(\frac{z+\zb}{2}\right)^2$ above, which comes from the $g_{\phi\phi}$ part of the metric, needs to be converted to Rindler using the corresponding coordinate transformation.

Already, given \eqref{eq:TBOPE2}, it is clear how the CFT$_2$ OPE blocks will look if we replace the stress tensor by a higher spin current. In particular, if the relevant components are still either $zz\dots z$ or $\zb\zb\dots\zb$ (or their projections). It is clear that the trace part will again be zero, as they will involve either metric factors $g_{zz}$ or $g_{\zb\zb}$, and we will ultimately get $\ell_B$ powers of $Z_z$ or $Z_{\zb}$. Here we just write down the final answer which takes the form
\begin{equation}\label{eq:BJOO3d}
	\mathcal{B}_{J_{zz\dots z}\mathcal{O}\mathcal{O}}=\tilde{C}_{d=2}\,\int_{D}\sqrt{-g^R}\,d\rho\,d\chi\,e^{\rho\,\ell_B}\,e^{\chi\,\Delta_{TB}}\,\mathcal{O}_{\Delta,\ell=0}(\rho,\chi,\phi)\,,
\end{equation}
and
\begin{equation}
	\mathcal{B}_{J_{\zb\zb\dots \zb}\mathcal{O}\mathcal{O}}=\tilde{C}_{d=2}\,\int_{D}\sqrt{-g^R}\,d\rho\,d\chi\,e^{-\rho\,\ell_B}\,e^{\chi\,\Delta_{TB}}\,\mathcal{O}_{\Delta,\ell=0}(\rho,\chi,\phi)\,.
\end{equation}
The most plausible and simplest possible generalization of these results to higher dimensions (only for all $z$ or all $\zb$ components) seem to be 
\begin{equation}\label{eq:zzopebhshd}
	\mathcal{B}^{k}_{J_{zz\dots z}\mathcal{O}\mathcal{O}}=\tilde{C}_d\,\int_{D}\sqrt{-g^R}\,d\rho\,d\chi\,e^{\rho\,\ell_{BT}}\,e^{\chi\,\Delta_{TB}}\left[\int\,d\Omega_{d-2}\right]\,\mathcal{O}_{\Delta,\ell=0}(\rho,\chi,\phi)\,,
\end{equation}
and
\begin{equation}\label{eq:zbzbtopebhshd}
	\mathcal{B}^{k}_{J_{\zb\zb\dots\zb}\mathcal{O}\mathcal{O}}=\tilde{C}_d\,\int_{D}\sqrt{-g^R}\,d\rho\,d\chi\,e^{-\rho\,\ell_{BT}}\,e^{\chi\,\Delta_{TB}}\left[\int\,d\Omega_{d-2}\right]\,\mathcal{O}_{\Delta,\ell=0}(\rho,\chi,\phi)\,.
\end{equation}
Given these expressions, we can now compute their resulting commutators with the modular Hamiltonian $H_{mod}$ and modular translation generator $P_{mod}$. They are given (computed only for $H_{mod}$) in appendix \ref{subapp:HBcomm}. As mentioned in the beginning of this section, as these OPE blocks are no longer eigenmodes of modular Hamiltonian, we don't expect for them to have a straightforward relation with the bulk fields on RT surface, unlike the case for CFT$_2$.

\section{Results and outlook\label{sec:conclude}}

Here we shall briefly summarize some of our key findings, with a few future directional comments.

\emph{Generalized OPE Blocks in AdS/CFT :} The work revisits the study of OPE blocks in the AdS/CFT framework, analyzing their modular properties and their connection to geodesic bulk fields. Notably, it includes new results on OPE blocks for timelike-separated operators, especially in higher dimensions (along with in two-dimensions in sections \ref{sec:ope} and \ref{sec:opehkllmatch}) with external operators being of unequal spins. Our approach differs from previous approaches, used previously for higher dimensions. The use of unequal spins highlights some subtleties that help us make connections with the dS/CFT dictionary. 


\emph{Euclidean OPE blocks for dS bulk fields :} The study extends the concept of the holographic dictionary to de Sitter  spacetime (section \ref{sec:dSconn}), using OPE blocks to describe bulk scalar fields integrated over geodesic surfaces (the area of these surfaces compute the pseudo-entropy in dS). By developing Euclidean OPE blocks in momentum space and in the shadow operator representation (appendix \ref{sec:OPEdS}), we demonstrate their natural connection to the geodesic bulk field in dS flat slicing. Utilizing principal series-represented external operators in the OPE to obtain a Fourier representation of the AdS bulk field, provides an alternative approach to previous methods, such as the X-ray/Radon transform or inverse Mellin transformation.


\emph{Higher-Dimensional generalizations :} The study generalizes the analysis of OPE blocks to higher-dimensional setups, addressing the challenges posed by their modular properties and their interplay with bulk fields in complex geometries. As we explicitly show, because these OPE blocks fail to be the eigenmodes of the Modular Hamiltonian, their representation as geodesic bulk field becomes unavailable (section \ref{sec:higherd}). However, if we perform an analytic continuation of our de Sitter result to AdS (upon dropping the non-normalizable piece), we will recover a \emph{different} representation of AdS bulk fields in terms of Lorentzian OPE blocks. Such a result will be consistent with the prior findings in \cite{Chen:2019fvi}.

\emph{Gravitons in AdS/CFT :} The study extends OPE block analysis to higher-spin fields, specifically spin-2  gravitons. In holographic gauge of AdS$_3$, geodesic gravitons furnish a representation of the modular Hamiltonian (appendix \ref{app:spin2case}).


As for the future direction, it would be interesting to investigate the existence of a modular Hamiltonian-like operator in the dS/CFT context. In \cite{Das:2023yyl}, such a quantity was indeed introduced, which was dubbed as pseudo-modular Hamiltonian. Given a bulk field residing on the hyperbolic extremal surfaces \eqref{eq:hypextsurface} are invariant under the flow of pseudo-modular Hamiltonian, it seems that a JLMS-type \cite{Jafferis:2015del} relation should also hold here.

\bigskip
\goodbreak
\centerline{\bf Acknowledgements}
\noindent
We would like to thank Bartek Czech for an early collaboration on this project. His insightful comments and enthusiasm kick-started this investigation. We also thank Pratik Das, Dan Kabat and Gaurav Katoch for their valuable feedback on various stages of this manuscript. The work of MN is supported by DST Inspire grant and the work of DS is supported by DST-FIST grant SR/FST/PSI-225/2016 and SERB MATRICS grant MTR/2021/000168.

\appendix
\section{Matrices and modular flows in Rindler and Poincar\'e AdS$_{3}$\label{appendix:matrices}}

We will use this appendix to mostly straighten out our notations of coordinates for AdS$_3$/ CFT$_2$ and also the associated modular Hamiltonians for a Rindler subregion. 

In embedding coordinates, the AdS$_3$ metric is given by 
\begin{equation}\label{eq:embedads}
	ds^2=-dU^2-dV^2+dX^2+dY^2,
\end{equation}
along with the constraint
\begin{equation}
	-U^2-V^2+X^2+Y^2=-R_{AdS}^2.
\end{equation}
We can go to the Rindler patch
\begin{equation}\label{eq:rindadshkll}
	ds^2=-\frac{r^2-r_+^2}{R_{AdS}^2}dt_R^2+\frac{R_{AdS}^2}{r^2-r_+^2}dr^2+r^2d\tilde{\phi}^2
\end{equation}
by making the following coordinate transformations:
\begin{eqnarray}
	U=\frac{R_{AdS}r}{r_+}\,\cosh\frac{r_+\tilde{\phi}}{R_{AdS}},\quad V=R_{AdS}\sqrt{\frac{r^2}{r_+^2}-1}\, \sinh\frac{r_+t_R}{R_{AdS}^2},\nonumber\\
	X=R_{AdS}\sqrt{\frac{r^2}{r_+^2}-1} \,\cosh\frac{r_+t_R}{R_{AdS}^2},\quad Y=\frac{R_{AdS}r}{r_+}\,\sinh\frac{r_+\tilde{\phi}}{R_{AdS}}.
\end{eqnarray}
In these coordinates, $-\infty<t_R,\tilde{\phi}<\infty$ and $r_+<r<\infty$. $r_+$ is the Rindler horizon and the boundary is at $r\to \infty$. Upon further coordinate rescalings of \eqref{eq:rindadshkll}, namely defining
\begin{equation}
	\tilde{r}=\frac{r}{r_+},\quad \rho=\frac{r_+\tilde{\phi}}{R_{AdS}},\quad \chi=\frac{r_+ t_R}{R_{AdS}^2},
\end{equation}
we end up with the following metric\footnote{In \cite{Hamilton:2006az,Hamilton:2006fh}, these same $\rho,\chi$ coordinates were called $\hat{\phi}$ and $\hat{t}$ respectively.}
\begin{equation}\label{eq:appbulkcoord0}
	ds^2=R_{AdS}^2\left(-\left(\tilde{r}^2-1\right)d\chi^2+\tilde{r}^2d\rho^2+\frac{d\tilde{r}^2}{\tilde{r}^2-1}\right).
\end{equation}
We have used this metric in \eqref{eq:RHKLLinR} e.g. We can consider the boundary limit of this spacetime by taking $\tilde{r}\to\infty$, where the above metric boils down to a metric conformal to 
\begin{equation}
	ds^2|_{boundary}=-d\chi^2+d\rho^2\,.
\end{equation}
So, this will be the boundary limit of the Rindler patch.
\\

We will also use the AdS$_3$ Poincar\'{e} patch, which is given by
\begin{equation}\label{eq:pads}
	ds^2=\frac{R_{AdS}^2}{Z^2}\left(-dt^2+dx^2+dZ^2\right).
\end{equation}
In its boundary limit $Z\to 0$, we again have a conformally flat metric given by
\begin{equation}\label{eq:pb}
	ds^2|_{boundary}=-dt^2+dx^2\,.
\end{equation}
The two bulk patches \eqref{eq:appbulkcoord0} and \eqref{eq:pads} are related to each other by the following coordinate transformation
\begin{equation}\label{eq:genctbulk}
	t=\frac{R_{AdS}\,\sqrt{\tilde{r}^2-1}\,\sinh\chi}{\tilde{r}\cosh\rho+\sqrt{\tilde{r}^2-1}\cosh\chi},\quad x=\frac{R_{AdS}\,\tilde{r}\,\sinh\rho}{\tilde{r}\cosh\rho+\sqrt{\tilde{r}^2-1}\cosh\chi},\quad Z=\frac{R_{AdS}}{\tilde{r}\cosh\rho+\sqrt{\tilde{r}^2-1}\cosh\chi}.
\end{equation}
The Rindler patch also provides a natural coordinate to study the entanglement wedge (EW) of a given boundary subregion. For example, for an interval $x=[L,R]$ in the CFT$_2$ in Poincar\'{e} coordinates (at a fixed time slice of the boundary), the corresponding  $(\rho,\chi,\tilde{r})$ provide only the spacetime contained within the EW of that subregion in AdS$_3$. This time, the coordinate transformations are a suitable generalization of \eqref{eq:genctbulk}:
\begin{align}\label{eq:ct2}
	t=\frac{R-L}{2}\frac{\sqrt{\tilde{r}^2-1}\sinh\chi}{\tilde{r}\cosh\rho+\sqrt{\tilde{r}^2-1}\cosh\chi},&\quad x=R+\frac{L-R}{2}\left(1-\frac{\tilde{r}\sinh\rho}{\tilde{r}\cosh\rho+\sqrt{\tilde{r}^2-1}\cosh\chi}\right)\nonumber\\ 
	&\text{and}\quad Z=\frac{R-L}{2}\frac{1}{\tilde{r}\cosh\rho+\sqrt{\tilde{r}^2-1}\cosh\chi}.
\end{align}
Once again, the resulting metric is given by \eqref{eq:appbulkcoord0}, but this time they only capture the EW of the subregion. Once the coordinate transformation is performed, all factors of $L,R$ drop out and the left and right endpoints of the subregion gets mapped to $\rho=-\infty$ and $\rho=\infty$ respectively. Also, the top and the bottom tips of its causal diamond located at 
\begin{equation}\label{eq:topbottom}
	x=\frac{R+L}{2}\qquad\text{with}\qquad t=\frac{R-L}{2}\quad \text{and}\quad t=\frac{L-R}{2}\,,
\end{equation}
gets mapped to $\chi\to \infty$ and $\chi\to -\infty$ respectively. The coordinate transformation between just the boundary part of the metric is obtained by taking $\tilde{r}\to \infty$ limit of \eqref{eq:ct2}: 
\begin{equation}\label{eq:genctbdry}
	x=R+\frac{L-R}{2}\left(1-\frac{\sinh\rho}{(\cosh\rho+\cosh\chi)}\right),\quad t=\frac{R-L}{2}\frac{\sinh\chi}{(\cosh\rho+\cosh\chi)}.
\end{equation} 
So, once we perform \eqref{eq:genctbdry} to \eqref{eq:pb}, the resulting metric is
\begin{equation}\label{eq:conffact}
	ds^2|_{boundary}=\frac{(L-R)^2}{4(\cosh\rho+\cosh\chi)^2} (d\rho^2-d\chi^2)=\Omega^2 (d\rho^2-d\chi^2)=-dt^2+dx^2.
\end{equation}
The readers may be familiar with a more specialized version of this coordinate transformation where the subregion is $x=[-R,R]$, which can be obtained by substituting $L=-R$ in the above. 

Particularly for AdS$_3$/CFT$_2$ (a version of the following is also present in higher dimensions), it is often useful to go to boundary lightcone coordinates, which in Poincar\'{e} $(x,t)$ we define as $z=x-t$ and $\bar{z}=x+t$.\footnote{People often use different conventions for defining the lightcone coordinate. We have used this version, so that it matches with the usual lightcone coordinate's definition after a Euclideanization by $t\to -i\tau_{E}$, where $\tau_E$ is the Euclidean time.} In these coordinates, the left and right endpoints (at $t=0$ slice) have coordinate values 
\begin{equation}\label{eq:lrinz}
	(z_L,\zb_{L})\equiv(L,\bar{L})=(L,L)\qquad \text{and}\qquad (z_R,\zb_{R})\equiv(R,\bar{R})=(R,R)\,.
\end{equation}
On the other hand, the top and bottom points of a causal diamond have coordinates
\begin{equation}\label{eq:tbinz}
	(z_T,\zb_{T})\equiv(T,\bar{T})=(L,R)\qquad \text{and}\qquad (z_B,\zb_{B})\equiv(B,\bar{B})=(R,L)\,.
\end{equation}
Such lightcone coordinates can then be related to the Rindler light cone coordinates $(w,\bar{w})$ via 
\begin{equation}\label{eq:ct1}
	w=f(z)=\log\frac{z-L}{R-z} \quad \text{and}\quad \wb=f(\zb)=\log\frac{\zb-\lb}{\rb-\zb}
\end{equation}
for left-right endpoints, and via
\begin{equation}\label{eq:cttb}
	w=f(z)=\log\frac{z-T}{B-z}\quad\text{and}\quad \wb=f(\zb)=\log\frac{\zb-\bb}{\tb-\zb}
\end{equation}
for top-bottom endpoints. These are Rindler lightcone coordinates as they are related to the boundary $(\rho,\chi)$ coordinates via $w=\rho-\chi$ and $\bar{w}=\rho+\chi$.

In summary, given a boundary subregion $x\in [L,R]$, the corresponding bulk transformations \eqref{eq:ct2} map its causal diamond to $(\rho, \chi)$ coordinates, with the associated bulk entanglement wedge (EW) horizon located at $\tilde{r}=1$. 

\subsection{Modular flows}\label{subapp:hmod}

It is well-known that for our single interval case of $x\in [L,R]$, the resulting modular Hamiltonian is written as (see e.g. \cite{Cardy:2016fqc})\footnote{See e.g. \cite{Witten:2018zxz} for a brief review on the ideas and subtleties of modular Hamiltonians etc. and their role as a boost operator for Rindler. We will not deal with such subtleties here, which have to do with algebraic structure of quantum field theory.} 
\begin{equation}\label{eq:cardytonni}
	H_{mod}=2\pi\left(\int_{-\infty}^{\infty}dw\, T_{ww}+\int_{-\infty}^{\infty}d\wb\, T_{\wb\wb}\right)=2\pi\int_A\frac{T_{zz}(z)}{f'(z)}dz+2\pi\int_A\frac{T_{\zb\zb}(\zb)}{f'(\zb)}d\zb\,,
\end{equation}
where $f(z)$ and $f(\zb)$ are the conformal transformations that appear in \eqref{eq:ct1}. This gives rise to the following modular Hamiltonian in $(z,\zb)$ plane
\begin{align}\label{eq:modhamlr}
	H_{mod}&=2\pi\left(\int_{L}^{R}dz\, T_{zz}(z)\frac{(L-z)(R-z)}{L-R}+\int_{L}^{R}d\bar{z}\, T_{\zb\zb}(z)\frac{(L-\zb)(R-\zb)}{L-R}\right)\nonumber\\
	&=H_{mod,rm}+H_{mod,\ell m}\,.
\end{align}
The $rm$ $(\ell m)$ in the subscript of $H_{mod}$ above denote the right-moving (left-moving) (or (anti-) holomorphic) parts of the equation. Here, in going between $(w,\wb)$ and $(z,\zb)$ coordinates, we have neglected a Schwarzian term. This term is relevant when computing the entanglement entropy as it gives the area piece, but at the level of modular Hamiltonian operator it is simply an irrelevant shift. 
The modular Hamiltonian above generates a flow of the boundary operators in the direction of the conformal Killing vector 
\begin{equation}\label{eq:killflow}
	\partial_\chi=\frac{1}{(R-L)}\left[\left(\left(\frac{R-L}{2}\right)^2-x^2-t^2\right)\frac{\partial}{\partial t}-2\,x\,t\,\frac{\partial}{\partial x}\right].
\end{equation}
We can also talk about the antisymmetric combination of $H_{mod,rm}$ and $H_{mod,\ell m}$ mentioned above 
\begin{equation}\label{eq:hmodrhodef}
	P_{mod}=H_{mod,rm}-H_{mod,\ell m}=2\pi\left(\int_{-\infty}^{\infty}dw T_{ww}-\int_{-\infty}^{\infty}d\wb T_{\wb\wb}\right).
\end{equation}
Denoting it by $P_{mod}$ above is a reminder that it is generating a flow in the spatial $\rho$ direction and is therefore naturally identified with the modular \emph{momentum} generator. In other words, it generates a conformal Killing flow in the direction
\begin{equation}
	\partial_\rho=\frac{1}{(R-L)}\left[\left(\left(\frac{R-L}{2}\right)^2-x^2-t^2\right)\frac{\partial}{\partial x}-2\,x\,t\,\frac{\partial}{\partial t}\right].
\end{equation}
A related discussion of these operators also appear in \cite{Czech:2017zfq}. 

For timelike intervals, for a discussion on a similar `timelike' modular Hamiltonian, see \cite{Das:2023yyl}. One can also define a modular Hamiltonian like quantity for the Euclidean case (relevant for de Sitter), as has also been studied there.

\section{$\mathcal{B}_{\mathcal{O}\mathcal{O}\mathcal{O}_{\Delta}}$ and $\mathcal{B}_{T_{\mu\nu}\mathcal{O}\mathcal{O}_\Delta}$ as eigenmodes of modular Hamiltonian}\label{app:HBcomm}

In this section, we will provide a derivation of \eqref{eq:HmodBcomm} and \eqref{eq:HBcomm2}, the fact that they are particular eigenmodes of boundary generators $H_{mod}$ and $P_{mod}$.  This is true for external operators with arbitrary spin, although below we will note the results for the cases when one of the external operators is a stress-tensor (the other being a scalar). At the end, we will also briefly comment about their higher dimensional generalizations. This is an extension of \cite{Das:2019iit}, which computed the scalar channel OPE blocks with left-right scalar external operators.

Below we will note down the result for the commutator of $H_{mod}$ and  $P_{mod}$ with $\mathcal{B}_{T_{zz}\mathcal{O}\mathcal{O}}$ giving rise to \eqref{eq:HmodBcomm} and \eqref{eq:HBcomm2}. For example, we recover \eqref{eq:HmodBcomm} when we put the spin of the bottom external operator to 0. Using the commutator between modular Hamiltonian and $\mathcal{O}$ \cite{Kabat:2017mun} (which follows from \eqref{eq:tzcomm})
\begin{equation}
	\left[H_{mod,rm}, \mathcal{O}(z,\zb)\right]=\Theta((z-T)(B-z))\frac{2\pi i}{B-T}\left[h(B+T-2z)+(z-T)(B-z)\partial_z\right]\mathcal{O}(z,\zb)\,,
\end{equation}
and \eqref{eq:opebtootb}, we obtain 
\begin{equation}
	\left[H_{mod,rm}, \mathcal{B}_{T_{zz}\mathcal{O}\mathcal{O}}\right]=2\pi i \left(\frac{\Delta_{TB}-\ell_{BT}}{2}\right)\mathcal{B}_{T_{zz}\mathcal{O}\mathcal{O}}\,.
\end{equation}
Similarly for the left moving part, we have
\begin{equation}
	\left[H_{mod,lm}, \mathcal{B}_{T_{zz}\mathcal{O}\mathcal{O}}\right]=2\pi i \left(\frac{\Delta_{TB}+\ell_{BT}}{2}\right)\mathcal{B}_{T_{zz}\mathcal{O}\mathcal{O}}\,.
\end{equation}
As a result, we have 
\begin{equation}\label{eq:HmodBTOOcomm2d}
	\left[H_{mod}, \mathcal{B}_{T_{zz}\mathcal{O}\mathcal{O}}\right]=2\pi i\, \Delta_{TB}\,\mathcal{B}_{T_{zz}\mathcal{O}\mathcal{O}}\,,
\end{equation}
which is essentially same as \eqref{eq:HmodBcomm} for the external scalar case. Taking the anti-symmetric combination, we then obtain 
\begin{equation}
	\left[P_{mod}, \mathcal{B}_{T_{zz}\mathcal{O}\mathcal{O}}\right]=-2\pi i\, \ell_{BT}\,\mathcal{B}_{T_{zz}\mathcal{O}\mathcal{O}}\,.
\end{equation}
For the anti-holomorphic components of the stress tensor, we obtain 
\begin{equation}
	\left[H_{mod,rm}, \mathcal{B}_{T_{\zb\zb}\mathcal{O}\mathcal{O}}\right]=2\pi i \left(\frac{\Delta_{TB}+\ell_{BT}}{2}\right)\mathcal{B}_{T_{\zb\zb}\mathcal{O}\mathcal{O}}
\end{equation}
and
\begin{equation}
	\left[H_{mod,lm}, \mathcal{B}_{T_{\zb\zb}\mathcal{O}\mathcal{O}}\right]=2\pi i \left(\frac{\Delta_{TB}-\ell_{BT}}{2}\right)\mathcal{B}_{T_{\zb\zb}\mathcal{O}\mathcal{O}}\,,
\end{equation}
which gives 
\begin{equation}
	\left[H_{mod}, \mathcal{B}_{T_{\zb\zb}\mathcal{O}\mathcal{O}}\right]=2\pi i\, \Delta_{TB}\,\mathcal{B}_{T_{\zb\zb}\mathcal{O}\mathcal{O}}\quad\text{and}\quad\left[P_{mod}, \mathcal{B}_{T_{\zb\zb}\mathcal{O}\mathcal{O}}\right]=2\pi i\, \ell_{BT}\,\mathcal{B}_{T_{\zb\zb}\mathcal{O}\mathcal{O}}\,.
\end{equation}
We see that the second equation above is essentially what we wrote more generally in \eqref{eq:HBcomm2}.

For completeness, let's also write down the left-right ($LR$) case results, which gives
\begin{align}
	&\left[H_{mod,rm}, \mathcal{B}_{T_{zz}\mathcal{O}\mathcal{O}}\right]=2\pi i \left(\frac{\ell_{LR}+\Delta_{LR}}{2}\right)\mathcal{B}_{T_{zz}\mathcal{O}\mathcal{O}}\,,\nonumber\\
	&\left[H_{mod,lm}, \mathcal{B}_{T_{zz}\mathcal{O}\mathcal{O}}\right]=2\pi i \left(\frac{\ell_{LR}-\Delta_{LR}}{2}\right)\mathcal{B}_{T_{zz}\mathcal{O}\mathcal{O}}\,,\nonumber\\
	&\left[H_{mod,rm}, \mathcal{B}_{T_{\zb\zb}\mathcal{O}\mathcal{O}}\right]=2\pi i \left(\frac{\Delta_{LR}-\ell_{LR}}{2}\right)\mathcal{B}_{T_{\zb\zb}\mathcal{O}\mathcal{O}}\quad\text{and}\nonumber\\
	& \left[H_{mod,lm}, \mathcal{B}_{T_{\zb\zb}\mathcal{O}\mathcal{O}}\right]=-2\pi i \left(\frac{\Delta_{LR}+\ell_{LR}}{2}\right)\mathcal{B}_{T_{\zb\zb}\mathcal{O}\mathcal{O}}\,.
\end{align} 
These ultimately lead to
\begin{equation}\label{eq:HtransBcomm}
	\left[H_{mod}, \mathcal{B}_{T_{zz}\mathcal{O}\mathcal{O}}\right]=2\pi i\, \ell_{LR}\,\mathcal{B}_{T_{zz}\mathcal{O}\mathcal{O}}\quad\text{and}\quad\left[P_{mod}, \mathcal{B}_{T_{zz}\mathcal{O}\mathcal{O}}\right]=2\pi i\, \Delta_{LR}\,\mathcal{B}_{T_{zz}\mathcal{O}\mathcal{O}}\,
\end{equation}
and
\begin{equation}
	\left[H_{mod}, \mathcal{B}_{T_{\zb\zb}\mathcal{O}\mathcal{O}}\right]=-2\pi i\, \ell_{LR}\,\mathcal{B}_{T_{\zb\zb}\mathcal{O}\mathcal{O}}\quad\text{and}\quad\left[P_{mod}, \mathcal{B}_{T_{\zb\zb}\mathcal{O}\mathcal{O}}\right]=2\pi i\, \Delta_{LR}\,\mathcal{B}_{T_{\zb\zb}\mathcal{O}\mathcal{O}}\,.
\end{equation}
We see that the second equation of \eqref{eq:HtransBcomm} is same as the scalar case  given in \eqref{eq:HBcomm}, and \eqref{eq:HmodBcomm2} is essentially the same as the first equation of \eqref{eq:HtransBcomm}.

\subsection{Modular flows of OPE blocks in arbitrary dimensions}\label{subapp:HBcomm}

Finally we move to higher dimensions. In this case $H_{mod,lm}$, $H_{mod,rm}$ and $H_{mod}$ all have expressions similar to \eqref{eq:modhamlr} for a subregion on the $t=0$ slice, where $z$ and $\zb$ have now been defined via \eqref{eq:higherdzzbardef}. In other words, we now have a $(z,\bar{z})$ plane as before and all the $\vec{\phi}$ directions span an orthogonal $S^{d-2}$ space. So, in a sense, we can effectively borrow our formalisms of two-dimensions with $\vec{\phi}$ being the spectator directions. In higher dimensions, the $[T,\mathcal{O}]$ commutator has an extra $(d-2)$-dimensional delta function over the angular direction, which is precisely needed to cancel out the extra $(d-2)$ integrals appearing in the definition of the higher dimensional $H_{mod}$, which generically takes the form \cite{Casini:2011kv} 
\begin{equation}
	H_{mod}=2\pi \int_{D}d^{d-1}x\, \frac{R^2-r^2}{2R}\,T_{00}(x)\,.
\end{equation}
However in this case, the OPE blocks such as $\mathcal{B}_{\mathcal{O}\mathcal{O}\mathcal{O}_{\Delta}}(T,B)$ are no longer eigenmodes of the modular Hamiltonian $H_{mod}$ (in fact, their eigenmode property  in CFT$_2$ can be traced back to their structure such as \eqref{eq:opetoolrrind} or \eqref{eq:opeeqk5} in Rindler coordinates).\footnote{It seems we can also define a quantity like $P_{mod}$ in higher dimensions, just as in two-dimensions.} However in higher dimensions, looking at equations such as \eqref{btoohighd}, \eqref{btbaroohighd}, \eqref{bttbaroohighd} or \eqref{btppoohighd}, we can see that due to the $\sqrt{-g^R}$ factor, the corresponding OPE blocks are no longer be eigenmodes of $H_{mod}$ or $H_{mod,\rho}$, even for $\mathcal{B}_{\mathcal{O}\mathcal{O}\mathcal{O}}$, $\mathcal{B}_{T_{zz}\mathcal{O}\mathcal{O}}$ or $\mathcal{B}_{T_{\zb\zb}\mathcal{O}\mathcal{O}}$. For completeness, we have given below the expressions of the resulting commutators 
\begin{align}
	&\left[H_{mod,lm},\mathcal{B}_{\mathcal{OOO}_{\Delta}} \right] = \pi i\Delta_{TB}\mathcal{B}_{\mathcal{OOO}_{\Delta}} +\tilde{C}_d \int_D\sqrt{-g(z)}\,dz\,d\bar{z}\left(\int d\Omega_{d-2}\right)\frac{2\pi i}{(\bar{T}-\bar{B})}\nonumber\\
	&\times \left\lbrace \frac{(\bar{T}-\bar{z})(\bar{z}-\bar{B})}{\bar{T}-\bar{B}}\right\rbrace^{\frac{\Delta-d}{2}}
	\left(\frac{\bar{z}-\bar{B}}{\bar{T}-\bar{z}} \right)^{\frac{\Delta_{TB}}{2}}\left\lbrace \frac{(z-T)(B-z)}{B-T}\right\rbrace^ {\frac{\Delta-d}{2}} \left(\dfrac{B-z}{z-T} \right)^ {\frac{\Delta_{TB}}{2}}\nonumber\\
	&\times\left\lbrace \left(\dfrac{2-d}{2} \right)(\bar{B}+\bar{T}-2\bar{z})+\frac{d-2}{2(z+\bar{z})}(\bar{T}-\bar{z})(\bar{z}-\bar{B}) \right\rbrace O_{\Delta}(z,\bar{z},\vec{\phi})
\end{align}
and 
\begin{align}
	&\left[H^{}_{mod,rm},\mathcal{B}_{\mathcal{OOO}_{\Delta}} \right] = \pi i\Delta_{TB}\mathcal{B}_{\mathcal{OOO}_{\Delta}}-\tilde{C}_d \int_D\sqrt{-g(z)}\,dz \,d\bar{z}\left(\int d\Omega_{d-2}\right)\frac{2\pi i}{(B-T)}\nonumber\\
	&\times \left\lbrace \frac{(\bar{T}-\bar{z})(\bar{z}-\bar{B})}{\bar{T}-\bar{B}}\right\rbrace^{\frac{\Delta-d}{2}}
	\left(\frac{\bar{z}-\bar{B}}{\bar{T}-\bar{z}} \right)^{\frac{\Delta_{TB}}{2}}\left\lbrace \frac{(z-T)(B-z)}{B-T}\right\rbrace^ {\frac{\Delta-d}{2}} \left(\dfrac{B-z}{z-T} \right)^ {\frac{\Delta_{TB}}{2}}\nonumber\\
	&\times\left\lbrace \left(\dfrac{2-d}{2} \right)({B}+{T}-2{z})+\frac{d-2}{2(z+\bar{z})}(z-T)(B-z) \right\rbrace O_{\Delta}(z,\bar{z},\vec{\phi})\,.
\end{align}
As a result, we have (an analogous expression is obtained for $P_{mod}$)
\begin{align}\label{eq:HBooohighd}
	&\left[H_{mod},\mathcal{B}_{\mathcal{OOO}_{\Delta}} \right] = 2 \pi i \,\Delta_{TB}\,\mathcal{B}_{\mathcal{OOO}_{\Delta}} +2\pi i\,\tilde{C}_d\left(\frac{d-2}{2}\right) \int_D\sqrt{-g(z)}\,dz \,d\bar{z}\,\left(\int d\Omega_{d-2}\right)\nonumber\\
	&\times \left\lbrace \frac{(\bar{T}-\bar{z})(\bar{z}-\bar{B})}{\bar{T}-\bar{B}}\right\rbrace^{\frac{\Delta-d}{2}}
	\left(\frac{\bar{z}-\bar{B}}{\bar{T}-\bar{z}} \right)^{\frac{\Delta_{TB}}{2}}\left\lbrace \frac{(z-T)(B-z)}{B-T}\right\rbrace^ {\frac{\Delta-d}{2}} \left(\dfrac{B-z}{z-T} \right)^ {\frac{\Delta_{TB}}{2}}\nonumber\\
	&\times\left\lbrace \left(\frac{B+T-2z}{B-T}\right)-\left(\frac{\bar{B}+\bar{T}-2\bar{z}}{\bar{T}-\bar{B}}\right)+\frac{1}{(z+\bar{z})}\left(\frac{(\bar{T}-\bar{z})(\bar{z}-\bar{B})}{\bar{T}-\bar{B}}-\frac{(z-T)(B-z)}{B-T}\right) \right\rbrace O_{\Delta}(z,\bar{z},\vec{\phi})\,.
\end{align}
We can also compute such higher dimensional commutators involving $\mathcal{B}_{T_{zz}\mathcal{OO}_{\Delta}}$ e.g., which gives
\begin{align}\label{eq:HBtoohighd}
	&\left[H_{mod},\mathcal{B}_{T_{zz}\mathcal{OO}_{\Delta}} \right] = 2\pi i\,\Delta_{TB}\mathcal{B}_{T_{zz}\mathcal{OO}_{\Delta}}+2\pi i\,\tilde{C}_d \left(\dfrac{d-2}{2} \right)\int_D\sqrt{-g(z)}dz \,d\bar{z}\,\left(\int d\Omega_{d-2}\right)\nonumber\\
	&\left\{\frac{(z-T)^{\frac{\Delta-d}{2}-\frac{\Delta_{TB}-l_B}{2}}(B-z)^{\frac{\Delta-d}{2}+\frac{\Delta_{TB}-l_B}{2}}}{(B-T)^{\frac{\Delta-d}{2}+l_B}}\right\}\times\left\{\frac{(\bar{T}-\bar{z})^{\frac{\Delta-d}{2}-\frac{\Delta_{TB}+l_B}{2}}(\bar{z}-\bar{B})^{\frac{\Delta-d}{2}+\frac{\Delta_{TB}+l_B}{2}}}{(\bar{T}-\bar{B})^{\frac{\Delta-d}{2}}}\right\}\nonumber\\
	&\times\left\{ \frac{(B+T-2z)}{B-T}-\frac{(\bar{B}+\bar{T}-2\bar{z})}{\bar{T}-\bar{B}}
	+\frac{1}{(z+\bar{z})}\left(\frac{(\bar{T}-\bar{z})(\bar{z}-\bar{B})}{\bar{T}-\bar{B}}-\frac{(z-T)(B-z)}{B-T}\right) \right\} O_{\Delta}(z,\bar{z},\vec{\phi})\,.
\end{align}
From \eqref{eq:HBooohighd} or \eqref{eq:HBtoohighd} we see that the second term above goes away completely for $d=2$, and in that case we recover the eigenmode equations \eqref{eq:HmodBcomm} or \eqref{eq:HmodBTOOcomm2d} respectively.
This failure of being an eigenmode is reflected in the fact that there doesn't anymore exist a straightforward relation between the OPE blocks and the bulk fields (at RT surfaces) in higher dimensions. Although OPE blocks are no longer eigenmodes, the bulk field on the RT surface still remain an eigenmode of the modular Hamiltonian (with zero eigenvalue). 

\section{Bulk gravitons, OPE blocks as modular Hamiltonians and their connections}\label{app:spin2case}

\subsection{Free gravitons in AdS$_3$}\label{subsec:RHKLLgrav}

In this section, we turn to the reconstruction of bulk gravitons in AdS$_3$/CFT$_2$. Our main motivation stems from the anticipation that the resulting bulk field located at the horizon or RT surface may be connected to the corresponding OPE block in stress-tensor channel, which is nothing but the modular Hamiltonian in CFT$_2$. Note that the locality issues are much more subtle for gauge fields such as spin-2 gravitons, as the Gauss law forbids the gauge fields to be localized at a given region of spacetime. This feature was studied and discussed in detail in \cite{Heemskerk:2012np,Kabat:2012hp,Kabat:2012av,Kabat:2013wga,Donnelly:2015hta,Donnelly:2016rvo,Kabat:2020nvj} to name a few. In particular, \cite{Kabat:2012hp} provided the free graviton dictionary in the so-called holographic or radial gauge, which puts all the radial components of the bulk graviton to zero. For an AdS$_{d+1}$ metric given by 
\begin{equation}\label{eq:padsfg}
	ds^2=\frac{R_{AdS}^2}{Z^2}\left(dZ^2+g_{\mu\nu}\,dx^\mu\,dx^{\nu}\right),\quad\text{with}\quad g_{\mu\nu}=\eta_{\mu\nu}+\frac{Z^2	}{R_{AdS}^2}h_{\mu\nu}\,,
\end{equation}
it was shown that $Z^2h_{\mu\nu}$ satisfies a smearing function relation equivalent of a massless scalar. As a result, the graviton smearing function takes the form
\begin{equation}\label{eq:klrsgrav}
	Z^2\,h_{\mu\nu}=\frac{1}{\text{vol}(B^d)}\int_{t^2+|y'|^2<Z^2}dt'\,d^{d-1}y'\,T_{\mu\nu}(t+t',x+iy')\,.
\end{equation}
Here the integral is over a $d$-dimensional ball $B^d$. As was also shown, for AdS$_3$ it takes particularly simple form of 
\begin{equation}\label{eq:ads3gravklrs}
	h_{zz}=T_{zz}\,,\qquad h_{\zb\zb}=T_{\zb\zb}\,,\qquad\text{and}\qquad h_{z\zb}=0\,.
\end{equation}

\subsection{OPE blocks as modular Hamiltonians}\label{subsec:OPEisH}

Here we start by showing that the OPE block in the stress tensor channel with two identical external operators is nothing but the modular Hamiltonian. This result was already derived in \cite{Czech:2016xec} which utilizes the $SO(2,2)$ symmetry structure of AdS$_3$ and the fact that the metric on the kinematic space of AdS$_3$ breaks down into dS$_2\times$dS$_2$. We here implement the shadow operator method instead. We will utilize the fact that the shadow operator of the stress-tensor is a spin $\tilde{\ell}=2$ operator of conformal dimension $\tilde{\Delta}=0$.
In other words, we have 
\begin{align}\label{eq:opebottlr}
	\mathcal{B}_{\mathcal{O}\mathcal{O}T}(L,R)&=C_{T} \int_{D_{z\zb}} d^2x\, \left\langle \mathcal{O}_{\Delta_{R}}(R,\bar{R})\, \mathcal{O}_{\Delta_{L}}(L,\bar{L})\, \tilde{T}^{\mu\nu}_{\tilde{\Delta}=0,\, \tilde{\ell}=2}(x)\right\rangle\, T_{\mu\nu}(x)\nonumber\\
	&=C_{T} \int_{D_{z\zb}} dz\, \left\langle \mathcal{O}_{\Delta_{R}}(R,\bar{R})\, \mathcal{O}_{\Delta_{L}}(L,\bar{L})\, \tilde{T}^{zz}_{\tilde{\Delta}=0,\, \tilde{\ell}=2}(z)\right\rangle\, T_{zz}(z)\nonumber\\
	&+C_{T} \int_{D_{z\zb}} \, d\bar{z}\, \left\langle \mathcal{O}_{\Delta_{R}}(R,\bar{R})\, \mathcal{O}_{\Delta_{L}}(L,\bar{L})\, \tilde{T}^{\zb\zb}_{\tilde{\Delta}=0,\, \tilde{\ell}=2}(z)\right\rangle\, T_{\zb\zb}(\zb)\nonumber\\
	&=\mathcal{B}_{\mathcal{O}\mathcal{O}T_{zz}}+\mathcal{B}_{\mathcal{O}\mathcal{O}T_{\zb\zb}}\nonumber\\
	&={C}_{T}\int_{D_{z\zb}}dz\, \left(\frac{(R-z)(z-L)}{R-L}\right)\,T_{zz}(z)+{C}_{T}\int_{D_{z\zb}}\,d\zb\,\left(\frac{(\zb-\bar{L})(\bar{R}-\zb)}{\bar{R}-\bar{L}}\right)\,T_{\zb\zb}(\zb)\,,
\end{align}
where in the last line we have used the structure of $\langle\mathcal{O}\mathcal{O}{T}_{zz}\rangle$ and $\langle\mathcal{O}\mathcal{O}{T}_{\zb\zb}\rangle$ correlator of CFT$_2$ and $C_T=\frac{\Gamma(4)}{(\Gamma(2))^2}=6$ (see also footnote \ref{fn:coeff}). Using $\Delta_{LR}=0$, we find both the right and left-moving parts of the modular Hamiltonian \eqref{eq:modhamlr}. Therefore we arrive at the known result \cite{Czech:2016xec}
\begin{equation}\label{eq:BHmodrel}
	\mathcal{B}_{\mathcal{O}\mathcal{O}T_{zz}}(L,R)+\mathcal{B}_{\mathcal{O}\mathcal{O}T_{\zb\zb}}(L,R)=\tilde{C} H_{mod}\,,
\end{equation}
where $\tilde{C}=\frac{C_T}{2\pi}=\frac{3}{\pi}$.\footnote{Note that if we evaluate the Euclidean OPE block in this manner, we can derive the expression of pseudo-modular Hamiltonian. See equation (101) of \cite{Das:2023yyl} for the latter.} Given the remarkably simple relation for three-dimensional free graviton correspondence \eqref{eq:ads3gravklrs}, it is now quite straightforward to find the connection between the corresponding OPE blocks and bulk gravitons. Given \eqref{eq:BHmodrel}, we can straightaway write down
\begin{align}
	\tilde{C} H_{mod}&=\mathcal{B}_{\mathcal{O}\mathcal{O}T_{zz}}(L,R)+\mathcal{B}_{\mathcal{O}\mathcal{O}T_{\zb\zb}}(L,R)\nonumber\\
	&=2\pi\tilde{C}\left(\int_{L}^{R}dz\, T_{zz}(z)\frac{(L-z)(R-z)}{L-R}+\int_{L}^{R}d\bar{z}\, T_{\zb\zb}(z)\frac{(L-\zb)(R-\zb)}{L-R}\right)\nonumber\\
	&=2\pi\tilde{C}\left(\int_{L}^{R}dz\, h_{zz}(z)\frac{(L-z)(R-z)}{L-R}+\int_{L}^{R}d\bar{z}\, h_{\zb\zb}(z)\frac{(L-\zb)(R-\zb)}{L-R}\right)\nonumber\\
	&=2\pi\tilde{C}\left(\int_{D}dw\, h_{ww}(w)+\int_{D}d\bar{w}\, h_{\wb\wb}(\wb)\right)\,.
\end{align}
The above simplified relation can be traced back to the fact that for AdS$_3$, the bulk graviton fluctuations have no radial dependence (which is related to the fact that AdS$_3$ has no propagating gravitons).
This scenario will of course change in higher dimensions. It will be an interesting check whether this is consistent with how gravitons flow under the modular Hamiltonian \cite{Kabat:2018smf}. 

\section{Shadow formalism in momentum space}\label{sec:OPEdS}

Here for completeness we discuss the shadow operator formalism for OPE blocks in momentum space. The Lorentzian formulation of the same has been discussed in \cite{Chen:2019fvi} where the three point functions are given by Wightman functions. On the other hand, here we discuss the same for Euclidean signatures. We can do a Fourier transformation to represent the CFT states in momentum eigenkets as 
\begin{equation}
	|\mathcal{O}_{\Delta,J}(k,z)\rangle= \int d^dx\,e^{ik.x}\,	|\mathcal{O}_{\Delta,J}(x,z)\rangle
\end{equation}
with the states satisfying the norm condition ($z$ are the polarization vectors)
\begin{equation}\label{eq:momentumnorm}
	\left\langle\mathcal{O}_{\Delta, J}\left(k_1, z_1\right) \mid \mathcal{O}_{\Delta, J}\left(k_2, z_2\right)\right\rangle=(2 \pi)^d \delta^{(d)}\left(k_1+k_2\right) E_{\Delta, J}\left(k_2 ; z_1, z_2\right) .
\end{equation}
where $E_{\Delta, J}(k)$ is the momentum two point function in Fourier space. Then the shadow operator state can be defined as
\begin{equation}
	| \tilde{\mathcal{O}}_{\bar{\Delta}, J}(k)\rangle = E_{\bar{\Delta}, J}(k)|\mathcal{O}_{\Delta, J}(k)\rangle.
\end{equation}
Using the constraint that two consecutive shadow operation on a CFT primary $\mathcal{O}(k)$ should bring it back to the original operator, we obtain
\begin{equation}
	\tilde{ \tilde{\mathcal{O}}}_{\Delta, J}(k)=E_{\Delta, J}(k)E_{\Bar{\Delta}, J}(k)\mathcal{O}_{\Delta, J}(k)
\end{equation}
and hence we can define the shadow two point function in momentum space as the inverse of $E_{\Delta, J}(k)$. With this, the resulting shadow states follows the norm definition (see footnote \ref{foot:alphadef} for the definition of $\alpha$)
\begin{equation}
	\langle \tilde{\mathcal{O}}_{\Bar{\Delta}, J}\left(k_1, z_1\right) | \tilde{\mathcal{O}}_{\Bar{\Delta}, J}\left(k_2, z_2\right)\rangle=(2 \pi)^d \delta^{(d)}\left(k_1+k_2\right)\alpha_{\Delta, J} \alpha_{\bar{\Delta}, J} E_{\Bar{\Delta}, J}\left(k_2 ; z_1, z_2\right) .
\end{equation}
Finally, the shadow projector can then be written as
\begin{equation}\label{eq:shadowprojector}
	|\mathcal{O}_{\Delta, J}|=\frac{1}{\alpha_{\Delta, J} \alpha_{\bar{\Delta}, J}} \int d^d k\,|\tilde{\mathcal{O}}_{\bar{\Delta}, J}(-k)\rangle\langle\mathcal{O}_{\Delta, J}(k)|.
\end{equation}
We could verify the validity of the above definition by plugging \eqref{eq:shadowprojector} in \eqref{eq:momentumnorm}.
\subsection{Embedding of the de Sitter flat patch}\label{subsec:dSembed}
The de Sitter hyperboloid in $\mathbb{R}^{1,d+1}$ is given by
\begin{equation}
X^2=-(X^{-1})^2 + (X^d)^2+\delta_{mn}X^mX^n=R_{dS}^2\qquad m=0,...,d-1
\end{equation} 
In the light cone coordinates combining $X^{-1}$ and $X^d$ we can embed the Lorentzian flat patch on the dS$_{2+1}$ hyperboloid via the following mapping 
\begin{equation}
\{X^+~,~X^-~,~X^m\}=\frac{R_{dS}}{\eta}\{1~,~x^2-\eta^2~,~x^m\}\qquad\text{with}\quad m=0,1
\end{equation}
and thus
\begin{align}
ds^2&=-dX^+dX^-+\delta_{mn}dX^mdX^n\nonumber\\
&=\frac{R_{dS}^2}{\eta^2}\left(-d\eta^2+(dx^0)^2+(dx^1)^2\right)
\end{align}
\subsection{Useful parametrization and identities}\label{sec:identities}

In order to obtain the momentum three-point function from the position three-point function we used the following Swinger parametrization from \cite{Chen:2019fvi} in right hand side of \eqref{eq:bessel}
\begin{align} 
	\int d^d x_3\, & e^{\mathrm{i} p \cdot x}\left(x_{13}^2\right)^{-\delta_1}\left(x_{23}^2\right)^{-\delta_2}=\frac{2 \pi^{\bar{d}}}{\Gamma\left(\delta_1\right) \Gamma\left(\delta_2\right)}\left(\frac{p^2}{4 x_{12}^2}\right)^{\frac{\delta_1+\delta_2-\bar{d}}{2}}\nonumber \\ 
	& \times \int_0^1 ~d\, u u^{\frac{\delta_1-\delta_2+\bar{d}}{2}-1}(1-u)^{\frac{\delta_2-\delta_1+\bar{d}}{2}-1} e^{\mathrm{i} p \cdot\left(u x_1+(1-u) x_2\right)} K_{\delta_1+\delta_2-\bar{d}}\left(\sqrt{u(1-u) p^2 x_{12}^2}\right)\,.
\end{align}
Also the differential operator of the polarization vectors introduced in \eqref{eq:bessel} is given by
\begin{align} 
	\mathscr{D}_J(\alpha, \lambda ; \beta, \mu) & = J !(\lambda+\mu)^J P_J^{(\alpha-1, \beta-1)}\left(\frac{\mu-\lambda}{\mu+\lambda}\right)\nonumber \\ 
	& =\sum_{r=0}^J\left(\begin{array}{l}J\nonumber \\ 
		r\end{array}\right)(\alpha+r)_{J-r}(\beta+J-r)_r(-\lambda)^r \mu^{J-r} \\ 
	& =\frac{(\alpha)_J}{2^J} \sum_{r=0}^J \frac{(-J)_r(\alpha+\beta+J-1)_r}{(\alpha)_r r !}{ }_2 F_1(1-\beta-J, r-J ; \alpha+r ;-1)(\lambda-\mu)^r(\lambda+\mu)^{J-r}\,,
\end{align}
where $P_J^{(\alpha,\beta)}$ are Jacobi polynomials of degree $J$ and $_2 F_1$ are Gauss Hypergeometric functions. The $\mathcal{Q}$-kernel is defined as
\begin{align}\label{eq:Qdef}
	\mathcal{Q}_{\Delta_1,\Delta_2,\Delta,J}\left(x_1,x_2,-p,z_3\right) &=\frac{1}{\left(x_{12}^2\right)^{\frac{\Delta_{12}^{+}}{2}-\tau}}\mathscr{D}_J \left(\delta_{12}^{+},z_3 \cdot \partial_{1} ;\delta_{12}^{-},z_3 \cdot \partial_{2}\right)\left(\frac{p^2}{4{x_{12}^{2}}}\right)^{\frac{\tau-{\bar{d}}}{2}}\nonumber \\
	&\times\int_0^1 du\,u^{\frac{\Delta_{12}^{+}-{\bar{d}}}{2}-1} \left(1-u\right)^{\frac{{\bar{d}}-\Delta_{12}^{-}}{2}-1} e^{ip \cdot x\left(u\right)}I_{\left({\bar{d}}-\tau\right)}\left(\sqrt{u\left(1-u\right)p^2x_{12}^2}\right).
\end{align}
Furthermore, we have also used the following identity in section \ref{sec:dSconn}
\begin{align} \label{eq:id1}
	&\left(x_{12}^2\right)^{\frac{\tau}{2}} \mathcal{D}_J(\left.\delta_{12}^{+}, z \cdot \partial_1 ; \delta_{12}^{-}, z \cdot \partial_2\right)\left(\frac{p^2}{4 x_{12}^2}\right)^{\frac{\tau-\bar{d}}{2}}\int_0^1 ~d u~ u^{\frac{\Delta_{12}^{-}+{\bar{d}}}{2}-1}(1-u)^{\frac{{\bar{d}}-\Delta_{12}^{-}}{2}-1} e^{\mathrm{i} p \cdot x(u)} I_{\tau-{\bar{d}}}\left(\sqrt{u(1-u) p^2 x_{12}^2}\right)\nonumber \\ &=\kappa_{\Delta, J}\left(x_{12}^2\right)^{\frac{\bar{\tau}}{2}} \frac{1}{J !({\bar{d}}-1)_J} E_{\Delta, J}\left(p ; z, d_{z^{\prime}}\right) \mathcal{D}_J\left(\bar{\delta}_{12}^{+}, z^{\prime} \cdot \partial_1 ; \bar{\delta}_{12}^{-}, z^{\prime} \cdot \partial_2\right)\left(\frac{p^2}{4 x_{12}^2}\right)^{\frac{\bar{\tau}-{\bar{d}}}{2}}\int_0^1 ~d u u^{\frac{\Delta_{12}^{-}+{\bar{d}}}{2}-1}\nonumber \\ 
	&\qquad\qquad\qquad\qquad\qquad\qquad\qquad\qquad\qquad\qquad\times(1-u)^{\frac{{\bar{d}}-\Delta_{12}^{-}}{2}-1} \,e^{\mathrm{i} p \cdot x(u)}~ I_{{\bar{d}}-\bar{\tau}}\left(\sqrt{u(1-u) p^2 x_{12}^2}\right)\,,
\end{align}
where we introduced $\kappa_{\Delta, J} = \frac{\Gamma(\Delta+J)}{\pi^{{\bar{d}}}\,(d-\Delta-1)_J\,\Gamma({\bar{d}}-\Delta)}\ $ and  $E_{\Delta,J}(-p)=\alpha_{(\Delta,J)}(p^2)^{\Delta-{\bar{d}}}$.

\bibliographystyle{utphys}
\bibliography{refs}
\end{document}